\documentclass[12pt]{article}

\usepackage{latexsym}
\usepackage{amssymb,amsfonts,amsmath}
\usepackage{graphicx} 
\usepackage{indentfirst}
\usepackage{bbm}
\usepackage{amssymb}
\usepackage{verbatim}
\usepackage{amsmath, amsthm,amssymb}
\usepackage{mathrsfs}
\usepackage{hyperref}
\usepackage{amsfonts}
\usepackage{dsfont}
\usepackage{cite}
\parskip=\medskipamount

\newcommand {\cA}{{\cal A}}
\newcommand {\cB}{{\cal B}}

\newcommand {\cD}{{\cal D}}
\newcommand {\cE}{{\cal E}}

\newcommand {\cL}{{\cal L}}
\newcommand {\cM}{{\cal M}}
\newcommand {\cN}{{\cal N}}

\newcommand {\cS}{{\cal S}}

\newcommand {\cV}{{\cal V}}
\newcommand {\cW}{{\cal W}}

\def\a{\alpha}

\def\b{\beta}

\def\d{\delta}
\def\e{\epsilon}
\def\f{\phi}
\def\g{\gamma}

\def\k{\kappa}
\def\l{\lambda}
\def\m{\mu}

\def\o{\omega}

\def\q{\theta}
\def\r{\rho}
\def\s{\sigma}
\def\t{\tau}

\def\x{\xi}
\def\z{\zeta}

\def\F{\Phi}
\def\J{\Psi}
\def\L{\Lambda}
\def\O{\Omega}

\def\S{\Sigma}
\def\U{\Upsilon}

\def\tr{{\rm tr}}
\def\rd{{\rm d}}
\def\ri{{\rm i}}

\newcommand{\ve}{\varepsilon}                            %new
\newcommand{\cDB}{{\bar\cD}}                            %new

\newcommand{\pa}{\partial}                           %new
\newcommand{\hf}{\frac12}
\newcommand{\vf}{\varphi}
\newcommand{\be}{\begin{equation}}
\newcommand{\ee}{\end{equation}}
\newcommand{\bea}{\begin{eqnarray}}
\newcommand{\eea}{\end{eqnarray}}
\newcommand{\non}{\nonumber}
\newcommand{\1}{{\underline{1}}}
\newcommand{\2}{{\underline{2}}}

\newcommand{\bm}[1]{\mbox{\boldmath$#1$}}

\def\double #1{#1{\hbox{\kern-2pt $#1$}}}

\newif\ifdtup

\def\de{{\nabla}}                                         % del
\newcommand{\bsubeq}{\begin{subequations}}
\newcommand{\esubeq}{\end{subequations}}
\numberwithin{equation}{section}

\begin{document}

\begin{titlepage}
\begin{flushright}
May, 2019 \\
\end{flushright}
\vspace{5mm}

\begin{center}
{\Large \bf 
Field theories with (2,0) AdS supersymmetry \\
in  $\bm{{\cal N}=1}$ AdS superspace}
\\ 
\end{center}

\begin{center}

{\bf
Jessica Hutomo and Sergei M. Kuzenko} \\
\vspace{5mm}

\footnotesize{
{\it Department of Physics M013, The University of Western Australia\\
35 Stirling Highway, Crawley W.A. 6009, Australia}}  
~\\

\vspace{2mm}
~\\
\texttt{jessica.hutomo@research.uwa.edu.au, sergei.kuzenko@uwa.edu.au}\\
\vspace{2mm}

\end{center}

\begin{abstract}
\baselineskip=14pt
In three dimensions, it is known that field theories possessing  extended $(p,q)$
anti-de Sitter (AdS) supersymmetry with ${\cal N}=p+q \geq 3$ can be realised in (2,0) AdS superspace. Here we present a formalism to reduce every field theory with (2,0) 
AdS supersymmetry to ${\cal N}=1$ AdS superspace. 
As nontrivial examples, we consider 
supersymmetric nonlinear sigma models formulated in terms of ${\cal N}=2$ chiral 
and linear supermultiplets. 
The $(2,0) \to (1,0)$ AdS reduction technique is then applied 
to the off-shell massless higher-spin supermultiplets in 
(2,0) AdS superspace 
constructed in \cite{HK18}. As a result, for each superspin value $\hat s$,
integer $(\hat s= s)$ or half-integer $(\hat s= s+\hf)$,
with $s=1,2,\dots $, we obtain two off-shell formulations for a massless 
${\cal N}=1$ superspin-$\hat s$ multiplet in AdS${}_3$.
These models prove to be related to each other by a superfield Legendre transformation  in the flat superspace limit, but the duality is not lifted 
to the AdS case. Two out of the four series of $\cN=1$ supersymmetric higher-spin models thus derived  are new. The constructed massless ${\cal N}=1$ supersymmetric higher-spin actions in AdS${}_3$ are used to formulate 
(i) higher-spin supercurrent multiplets in $\cN=1$ AdS superspace; and
(ii) new  topologically massive higher-spin off-shell supermultiplets. 
Examples of ${\cal N}=1$ higher-spin supercurrents are given for models of a complex scalar supermultiplet. 
We also present two new off-shell formulations for a massive ${\cal N}=1$ gravitino supermultiplet in AdS${}_3$.
\end{abstract}
\vspace{5mm}

\vfill

\vfill
\end{titlepage}

\newpage
\renewcommand{\thefootnote}{\arabic{footnote}}
\setcounter{footnote}{0}

\tableofcontents{}
\vspace{1cm}
\bigskip\hrule

\allowdisplaybreaks

%%%%%%%%%%%%%%%%%%%%%%%%%%%%%%%%
%%%%%%%%%%%%%%%%%%%%%%%%%%%%%%%%

\section{Introduction}

In three spacetime dimensions, the AdS  group is a product of two simple groups, 
\bea
\rm SO(2,2) \cong \Big( SL(2, {\mathbb R}) \times SL( 2, {\mathbb R}) \Big)/{\mathbb Z}_2~,
\eea
and so are its supersymmetric extensions  
${\rm OSp} (p|2; {\mathbb R} ) \times  {\rm OSp} (q|2; {\mathbb R} )$.\footnote{More
general  AdS supergroups exist for $\cN\geq 4$.}
This implies that  $\cN$-extended AdS supergravity exists in several incarnations  \cite{AT}, which are known as the  $(p,q)$ AdS supergravity theories,
where the integers $p \geq q\geq 0$ are such that $\cN=p+q$.   
The so-called $(p,q)$ AdS superspace \cite{KLT-M12}
\bea
{\rm AdS}_{(3|p,q)} = \frac{ {\rm OSp} (p|2; {\mathbb R} ) \times  {\rm OSp} (q|2; {\mathbb R} ) } 
{ {\rm SL}( 2, {\mathbb R}) \times {\rm SO}(p) \times {\rm SO}(q)}
\label{1.2}
\eea
may be interpreted as  a maximally symmetric solution 
of  $(p,q)$ AdS supergravity.\footnote{In the case of $\cN=1$ AdS supersymmetry, both notations 
$(1,0)$ and $\cN=1$ are used in the literature. We will often use the notation 
AdS${}^{3|2}$ for $\cN=1$ AdS superspace.}
Within the off-shell formulation for $\cN$-extended conformal supergravity 
which was first sketched in \cite{HIPT} and then fully developed in \cite{KLT-M11},
${\rm AdS}_{(3|p,q)}$
originates as  a maximally symmetric supergeometry 
with covariantly constant torsion and curvature 
generated by a symmetric  torsion $S^{IJ}= S^{JI}$, with the structure-group indices $I, J$ 
taking values from 1 to $\cN$.  It turns out that $S^{IJ}$ is nonsingular
and can be brought to the form
\bea
S^{IJ}=\cS\,{\rm{diag}}(\,  \overbrace{+1,\cdots,+1}^{p} \, , \overbrace{-1,\cdots,-1}^{q=\cN-p} \,)
~,
\eea
for some positive parameter $\cS$ of unit dimension. 
For $p=\cN\geq 4$ and $q=0$, 
there exist more general AdS superspaces \cite{KLT-M12}
than the conformally flat ones defined by \eqref{1.2}.

In  the extended  $\cN=p+q \geq 3$ case,
general $(p,q)$ supersymmetric field theories in AdS${}_3$ can be realised
 in (2,0) AdS superspace, 
 ${\rm AdS}_{(3|2,0)}$ \cite{KLT-M12,BKT-M}.\footnote{General
aspects of (2,0) supersymmetric field theory in AdS${}_3$ were studied in \cite{KT-M11}.}
 Such realisations are often useful for applications, for instance,  
 in order to study the target space geometry of supersymmetric nonlinear 
 $\s$-models in AdS${}_3$ \cite{BKT-M}. It is worth elaborating on the $\s$-model story
 in some more detail.
 For the $\cN=3$ and $\cN=4$ choices, manifestly $(p,q)$ supersymmetric formulations
  have been constructed \cite{KLT-M12} for the most general nonlinear $\s$-models in AdS${}_3$
 (these formulations make use of the curved superspace techniques developed in \cite{KLT-M11}). 
 This manifestly supersymmetric setting is very powerful since it allows one to generate 
 arbitrary nonlinear $\s$-models with $(p,q)$ AdS supersymmetry. 
 However, it also has a drawback 
 that the hyperk\"ahler geometry of the $\s$-model target space is hidden. 
 In order to uncover this geometry, 
 the formulation of the nonlinear $\s$-model 
 in (2,0) AdS superspace becomes truly 
 indispensable \cite{BKT-M}.\footnote{Analogous results exist in four dimensions.
 The most general $\cN=2$ supersymmetric $\s$-model in 
$\rm AdS_4$ was constructed \cite{BKsigma1,BKsigma2}
using a formulation in terms of $\cN=1$ covariantly chiral superfields, 
as an extension of the earlier analysis in the super-Poincar\'e case
\cite{HullKLR,LR}.
One of the main virtues of the $\cN=1$ formulation 
\cite{BKsigma1,BKsigma2} is its geometric character, however the second supersymmetry is hidden. General off-shell $\cN=2$ supersymmetric $\sigma$-models in $\rm AdS_4$ were actually formulated a few years earlier \cite{KT-M-4D-2008}
in $\cN=2$ AdS superspace. The latter approach makes $\cN=2$ supersymmetry 
manifest, but the hyperk\"ahler geometry of the $\s$-model target space is hidden.
The two $\s$-model formulations are related via the $\cN=2 \to \cN=1$ AdS superspace reduction \cite{BKLT-M}.}
 
 This work is somewhat similar in spirit to \cite{BKT-M,BKLT-M}, 
 however our goals are quite different. 
 Specifically, we develop a formalism to reduce every field theory with (2,0) 
AdS supersymmetry to ${\cal N}=1$ AdS superspace. 
This formalism is then applied to carry out 
the $(2,0) \to (1,0)$ AdS reduction of the off-shell massless higher-spin 
supermultiplets in ${\rm AdS}_{(3|2,0)}$
constructed in \cite{HK18}. 
There are at least two motivations for pursuing such an application. 
Firstly,  certain theoretical arguments imply
that there exist more general off-shell massless higher-spin 
$\cN=1$ supermultiplets in AdS${}_3$ than those described in \cite{KP}. 
Secondly, $\cN=1$ supermultiplets of conserved higher-spin currents 
have never been constructed in AdS${}_3$ (except for the superconformal multiplets
of conserved currents  in Minkowski superspace \cite{NSU} 
which can readily be lifted to AdS${}_3$).
Both issues will be addressed below. In particular, we will derive
new off-shell higher-spin $\cN=1$ supermultiplets in AdS${}_3$, 
which will be used to construct 
new  topologically massive higher-spin  supermultiplets.

The table of contents reflects the structure of the paper. Our notation and conventions 
follow \cite{KLT-M11}.

%%%%%%%%%%%%%%%%%%%%%%%%%%%%
%%%%%%%%%%%%%%%%%%%%%%%%%%%%%%

\section{(2,0) $\to$ (1,0) AdS superspace reduction} \label{section2}

The aim of this section is to elaborate on the details of procedure for reducing 
the field theories  in  (2,0) AdS superspace to ${\cal N}=1$ AdS superspace. 
Explicit examples of such a reduction are given by considering supersymmetric 
nonlinear $\s$-models. 

%%%%%%%%%%%%%%%%%%%%%%%%%%%%%%%%
%%%%%%%%%%%%%%%%%%%%%%%%%%%%%%%%%

\subsection{Geometry of (2,0) AdS superspace: Complex basis}

We begin by briefly reviewing the key results
concerning  (2,0) AdS superspace, see \cite{KT-M11, BKT-M} for the details.
There are two ways to describe the geometry of (2,0) AdS superspace, 
which correspond to making use of either a real or a complex basis for the 
spinor covariant derivatives. We first consider the formulation in the complex basis.
 
The covariant derivatives of (2,0) AdS superspace are
\bea
\cD_{{\cA}}=(\cD_{{a}}, \cD_{{\a}},\bar \cD^\a)
=
E_{\cA}
+\O_{\cA}+\ri \F_{{\cA}} J~, \qquad E_\cA=E_\cA{}^\cM \frac{\pa}{\pa z^\cM}~.
\label{CovDev}
\eea
where $z^\cM = (x^m, \q^\m , \bar \q_\m)$ are local superspace coordinates,  and
$J$ is the generator of the $R$-symmetry group, ${\rm U(1)}_R$.
The generator $J$ is defined to act on the covariant derivatives as follows:
\bea
{[} J,\cD_{\a}{]}
=\cD_{\a}~,
\qquad
{[} J,\cDB^{\a}{]}
=-\cDB^\a~,
\qquad 
{[}J,\cD_a{]}=0~.
\eea
The Lorentz connection, $\O_\cA$, can be written in several equivalent forms, which are
\bea
\O_\cA=\hf\O_{\cA}{}^{bc} M_{bc}= -\O_{\cA}{}^b M_b
=\hf\O_{\cA}{}^{\b\g}M_{\b\g}~.
\eea
The relations between the Lorentz generators with two vector indices 
($M_{ab}= -M_{ba}$), one vector index ($M_a$)
and two spinor indices ($M_{\a\b} = M_{\b\a} $) 
are given in Appendix \ref{AppendixA}.

The covariant derivatives of (2,0) AdS superspace obey the following graded commutation relations:
\begin{subequations} \label{deriv20}
\bea
\{\cD_\a,\cD_\b\}&=&0~,\qquad
\{\bar \cD_\a,\bar \cD_\b\}=0~,\\
\{\cD_\a,\cDB_\b\}
&=&
-2 \ri \big( \cD_{\a\b} -2 \cS M_{\a \b} \big)
-4 \ri \ve_{\a\b} \cS J ~,\\
{[}\cD_{a},\cD_\b {]}
&=&(\g_a)_\b{}^\g \cS \cD_{\g}~, \quad
{[}\cD_{a},\cDB_\b{]}
= (\g_a)_\b{}^{\g}\cS \bar \cD_{\g}~, \\
{[}\cD_a,\cD_b]{}
&=&-4 \cS^2 M_{ab} ~.
\eea
\end{subequations}
Here the parameter $\mathcal{S}$ is related to the AdS scalar curvature as $R= -24 \mathcal{S}^2$.

There exists a universal formalism to determine isometries of curved superspace backgrounds
in diverse dimensions \cite{BK,K15}. This formalism was used in \cite{KT-M11} to 
compute the isometries of (2,0) AdS superspace
(as well as supersymmetric
backgrounds in off-shell $\cN=2$ supergravity theories \cite{KLRST-M}).
The isometries of (2,0) AdS superspace are generated by the Killing supervector fields 
$\z^\cA E_\cA$, which are defined to solve the master equation 
\begin{subequations}\label{Killing}
\bea
\big{[}\z+\hf l^{bc} M_{bc}+\ri \t J
\, ,\cD_\cA\big{]}=0~,
\eea
where 
\bea
\z= \z^\cB \cD_\cB =\z^b\cD_b+\z^\b\cD_\b+\bar \z_\b\bar \cD^\b~,
\qquad \overline{\z^b}=\z^b~,
\eea 
\end{subequations}
and $\t$ and $l^{bc}$ are some real U(1)${}_R$ and Lorentz superfield parameters, respectively.
It follows from eq. \eqref{Killing} that the parameters $\z_\a, \t$  and $l_{\a\b}$ are uniquely expressed in terms of the vector parameter $\z_{\a\b}$ as follows:
\bea
\z_\a= \frac{\ri}{6} \bar \cD^\b \z_{\b\a}~,\quad 
\t =\frac{\ri}{2} \cD^\a\z_\a 
~, \qquad 
l_{\a\b} =2\big( \cD_{(\a } \z_{\b)} -\cS \z_{\a\b} \big) ~.
\eea 
The vector parameter $\z_{\a\b}$ satisfies the equation 
\bea
\cD_{(\a}\z_{\b\g)}=0~.
\eea
This implies the standard Killing equation, 
\bea
\cD_a \z_b + \cD_b \z_a=0~.
\eea
One may also prove the following relations
\bea
 \bar \cD_\a \t
=\frac{\ri}{3}\bar \cD^\b l_{\a\b} = 4\cS \z_\a~,
\qquad 
%%%
\bar \cD_\a\z_\b=0~, \qquad 
\cD_{(\a} l_{\b\g)}=0~.
\eea
The Killing supervector fields prove to
generate the supergroup 
$\rm OSp(2|2;{\mathbb R}) \times Sp(2,{\mathbb R})$, 
the isometry group of (2,0) AdS superspace.
Rigid supersymmetric field theories  in (2,0) AdS superspace are required to be  invariant under the isometry transformations. An infinitesimal isometry transformation acts on a tensor superfield $\bm U$ (with suppressed indices) by the rule
\bea
\d_\z {\bm U} = \big(\z+\hf l^{bc} M_{bc}+\ri \t J
\big) {\bm U}~.
\eea

%%%%%%%%%%%%%%%%%%%%%%%%%%%%%%%%
%%%%%%%%%%%%%%%%%%%%%%%%%%%%%%%%

\subsection{Geometry of (2,0) AdS superspace: Real basis}

Instead of dealing with  the complex basis for the (2,0) AdS spinor covariant derivatives, eq. \eqref{CovDev}, 
it is more convenient to switch to a real basis in order 
to carry out reduction to $\cN=1$ AdS superspace ${\rm AdS}^{3|2}$.
Following \cite{KLT-M12}, such a  basis is introduced by replacing the complex operators $\cD_\a$ and $\bar \cD_\a$ with 
${\bm \de}_\a^I= ( {\bm \de}_\a^\1, {\bm \de}_\a^\2 )$ defined as follows:
 \bea
& \cD_\a=\frac{1}{\sqrt{2}}({\bm \nabla}_\a^{\1}-\ri {\bm \nabla}_\a^{\2})~,~~~
  \bar \cD_\a=-\frac{1}{\sqrt{2}}({\bm \nabla}_\a^{\1}+\ri {\bm \nabla}_\a^{\2})~.
  \label{N1-deriv}
 \eea
In a similar way we introduce real coordinates, $z^\cM = (x^m, \q^\m_I)$, 
to parametrise (2,0) AdS superspace. 
Defining ${\bm \nabla}_a =\cD_a$, the algebra of  (2,0) AdS covariant derivatives 
\eqref{deriv20} turns into\footnote{The antisymmetric tensors $\ve^{IJ}$ and $\ve_{IJ} $
are normalised as $\ve^{\1\2} = \ve_{\1\2} =1$.}
\bsubeq \label{2_0-alg-AdS}
\bea
&\{ {\bm \nabla}_\a^I, {\bm \nabla}_\b^J\}=
2\ri\d^{IJ} {\bm \nabla}_{\a\b}
-4\ri \d^{IJ} \cS M_{\a\b}
+4\ve_{\a\b}\ve^{IJ} \cS J
~,
\label{2_0-alg-AdS-1}
\\
%%%%%%%%%%%%
&{[} {\bm \nabla}_{a}, {\bm \nabla}_\b^J{]}
=
\cS (\g_a)_\b{}^\g {\bm \nabla}_{\g}^J
~, \qquad 
{[} {\bm \nabla}_{a}, {\bm \nabla}_b{]}
=
-4\cS^2 M_{ab}
~,
\label{2_0-alg-AdS-2}
\eea
\esubeq
The action of the ${\rm U(1)}_R$ generator on the spinor covariant derivatives 
is given by
\bea
{[}J, {\bm \nabla}_\a^I{]} = -\ri \ve_{I J}  {\bm \nabla}_{\a}^J~.
\eea

As may be seen from \eqref{2_0-alg-AdS},
the graded commutation relations for the operators $ {\bm \de}_a$ and 
$ {\bm \de}_\a^{\1}$
have the following properties: 
\begin{enumerate} 
\item These (anti-)commutation relations do not involve $ {\bm \de}_\a^{\2}$,
\bsubeq \label{2.144}
\bea
&\{  {\bm \nabla}_\a^\1,  {\bm \nabla}_\b^\1\}=
2\ri  {\bm \nabla}_{\a\b}
-4\ri  \cS M_{\a\b}
~,
\\
%%%%%%%%%%%%
&{[}  {\bm \nabla}_{a},  {\bm  \nabla}_\b^\1{]}
=
\cS (\g_a)_\b{}^\g  {\bm \nabla}_{\g}^\1
~,\qquad 
{[}  {\bm \nabla}_{a},  {\bm \nabla}_b{]}
=
-4\cS^2 M_{ab}
~.
\eea
\esubeq
\item
 Relations \eqref{2.144} are identical
to the algebra of the covariant derivatives of 
${\rm AdS}^{3|2}$, see \eqref{deriv20}.
\end{enumerate}
We thus see that ${\rm AdS}^{3|2}$ is naturally  embedded in
(2,0) AdS  superspace  as a subspace. The real Grassmann variables of (2,0) AdS superspace, 
$\q^\m_I =(\q^\m_{\1}, \q^\m_{\2} )$, may be chosen in such a way that 
${\rm AdS}^{3|2}$ corresponds to the surface defined by $\q^\m_{\2} =0$.
We also note that no U$(1)_R$ curvature  
is present in the algebra of $\cN=1$ AdS covariant derivatives.
These properties make possible a consistent  $(2,0)  \to (1,0)$
AdS superspace reduction.

Now we will recast  the fundamental properties of  
the (2,0) AdS Killing supervector fields in the real representation  \eqref{N1-deriv}.
The isometries of  (2,0) AdS superspace are described in terms of 
those first-order operators 
\begin{subequations}\label{2.155}
\bea
\z := \z^{\cB}  {\bm  \de}_{\cB} = \z^b  {\bm \de}_b+\z^\b_J  {\bm \de}_\b^J~, 
\qquad J= \1,\2~,
\label{realvect}
\eea
which solve the equation
\bea
\big{[}\z + \hf l^{bc} M_{bc} + \ri \t J ,  {\bm \de}_\cA\big{]}=0~, \label{Killing-real}
\eea
\end{subequations}
for some real parameters $\t$ and $l^{ab} = -l^{ba}$. 
Equation \eqref{Killing-real} is equivalent to
\begin{subequations} \label{K-eq-real1}
\bea
 {\bm \de}_{\a}^I \z_{\b}^J &=& -\ve_{\a \b} \ve^{IJ} \t + \cS \d^{IJ} \z_{\a \b}+ \hf \d^{IJ} l_{\a \b}~, \label{K-eq-real11}\\
 {\bm \de}_{\a}^I \z_{b} &=& 2 \ri \, \z^{\b I} (\g_b)_{\a \b}~,\\
 {\bm \de}_{\a}^I \t &=& -4 \ri\, \cS \ve^{IJ} \z_{\a J}~,\\
 {\bm \de}_{\a}^I l_{\b \g} &=& 8 \ri\, \cS \ve_{\a (\b} \z_{\g)}^I~, \label{K-eq-real14}
\eea
\end{subequations}
and
\begin{subequations} 
\bea
 {\bm \de}_{a} \z_{b} &=& l_{ab}= -l_{ba}~, \label{keq} \\
 {\bm \de}_{a} \z^{\b}_I &=& -\cS \z^{\a}_I (\g_a)_{\a}^{\,\b}~, \label{ks-eq} \\
 {\bm \de}_{a} \t &=& 0~,\\
 {\bm \de}_{a} l^{bc} &=& 4 \cS^2(\d_a^b \z^c - \d_a^c \z^b)~.
\eea
\end{subequations}
Some nontrivial implications of the above equations 
which will be important for our subsequent consideration are: 
\begin{subequations} \label{K-eq-real2}
\bea
 {\bm \de}^I_{(\a} \z_{\b \g)} &=&0 ~, \qquad   {\bm \de}^I_{(\a} l_{\b \g)} = 0~, \\
 {\bm \de}^{I}_{(\a} \z^{J}_{\b)} &=& 2 \cS \, \d^{IJ} \z_{\a \b}~,
\qquad
 {\bm \de}^{\g (I}\z^{J)}_{\g}=0~,\\
 \z^{I \a} &=& \frac{\ri}{6}  {\bm \de}^I_{\b} \z^{\a \b} 
 = \frac{\ri}{12 \cS}  {\bm \de}^I_{\b} l^{\a \b} 
= -\frac{\ri}{4 \cS} \ve^{IJ}  {\bm \de}_J^{\a}\, \t  ~,\\
\t &=& -\frac{1}{4} \ve_{IJ}  {\bm \de}^{\g I} \z^J_{\g}~.
\eea 
\end{subequations}
Equation \eqref{keq} implies that $\z_a$ is a Killing vector field,
\bea
 {\bm \de}_a \z_b +  {\bm \de}_b \z_a =0~,
\eea
while \eqref{ks-eq} is a Killing spinor equation. The real parameter $\t$ 
is constrained by 
\bea
( {\bm \de}^{\2})^2 \t = (  {\bm \de}^\1)^2 \t = 8\ri \cS \t~, \quad
{\bm \de}_a \t =0~.
\eea

%%%%%%%%%%%%%%%%%%%%%%%%%%%%%%%%%
%%%%%%%%%%%%%%%%%%%%%%%%%%%%%%%%%

\subsection{Reduction from (2,0) to ${\cN=1}$ AdS superspace}

Given a tensor superfield ${\bm U}(x, \q_I)$ on (2,0) AdS superspace, 
 its $\cN=1$ projection (or bar-projection) is defined by 
\bea
{\bm U}|:= {\bm U}(x, \q_I)|_{\q_{\2} =0}
\eea
in a {\it special coordinate system} to be specified below.
By definition, ${\bm U}|$ depends on 
the real  coordinates $z^M= (x^m, \q^\m)$, with $\q^\m:=\q^\m_{\1}$,  
which will be used to parametrise $\cN=1$ AdS superspace ${\rm AdS}^{3|2}$. 
For the (2,0) AdS covariant derivative
\bea
{\bm \nabla}_{\cA}=({\bm \nabla}_a , {\bm \nabla}_\a^I) 
=
E_\cA{}^\cM \frac{\pa}{\pa z^\cM}
+\hf\O_{\cA}{}^{bc} M_{bc}
+\ri \F_{{\cA}} J~,
\eea
its bar-projection is defined as
\bea
{\bm \nabla}_{\cA}|=E_\cA{}^\cM | \frac{\pa}{\pa z^\cM}
+\hf\O_{\cA}{}^{bc} | M_{bc}
+\ri \F_{{\cA}} |J~.
\eea

We use the freedom to perform general coordinate, local Lorentz and  U$(1)_R$ 
transformations  to choose the following  gauge condition
\bea
{\bm \nabla}_a |=\nabla_a~, \qquad {\bm \nabla}^\1_\a|=\nabla_\a~, 
\label{224}
\eea
where 
\bea
\nabla_A = (\nabla_a , \nabla_\a ) = E_A{}^M \frac{\pa}{\pa z^M}
+\hf\o_{A}{}^{bc} M_{bc}
\eea
denotes the set of covariant derivatives for ${\rm AdS}^{3|2}$, which obey 
 the following graded commutation relations: 
\bsubeq 
\bea
&\{  { \nabla}_\a,  { \nabla}_\b \}=
2\ri  { \nabla}_{\a\b}
-4\ri  \cS M_{\a\b}
~,
\\
%%%%%%%%%%%%
&{[}  { \nabla}_{a},  {  \nabla}_\b{]}
=
\cS (\g_a)_\b{}^\g  { \nabla}_{\g}
~,\qquad 
{[}  { \nabla}_{a},  { \nabla}_b{]}
=
-4\cS^2 M_{ab}
~.
\eea
\esubeq
In such a coordinate system,
the operator ${\bm \nabla}_\a^\1|$   contains no 
partial derivative with respect to $\q_\2$. As a consequence, 
 $\big({\bm \nabla}^\1_{{\a}_1} \cdots {\bm \nabla}^\1_{{\a}_k} {\bm U} \big)\big|
= \nabla_{{\a}_1} \cdots  \nabla_{{\a}_k} {\bm U}|$, for any positive integer $k$,  
where $\bm U$ is a tensor superfield on (2,0) AdS superspace.
Let us study how the $\cN=1$ descendants of $\bm U$ defined by
$U_{\a_1 \dots \a_k}:=
\big({\bm \nabla}^\2_{{\a}_1} \cdots {\bm \nabla}^\2_{{\a}_k} {\bm U} \big)\big|$
 transform under the (2,0) AdS isometries, 
 with $k$ a non-negative integer.

We introduce 
the $\cN=1$ projection of the (2,0) AdS Killing supervector field \eqref{2.155}
\bea
\z| = \x^b \de_b + \x^{\b} \de_{\b}+ \e^{\b} {\bm \de}_{\b}^{\2}|~,\qquad 
\x^b := \z^b|~, ~~\x^{\b}:= \z_{\1}^{\b}|~, ~~\e^{\b} :=  \z_{\2}^{\b}|~.
\eea
We also introduce the $\cN=1$ projections of the  
Lorentz and U$(1)_R$ parameters in \eqref{2.155}:
\bea
\l^{bc} := l^{bc}|~,~ \qquad \e := \t|~.
\eea
It follows from \eqref{2.155} that the $\cN=1$ parameters $\x^B = (\x^b, \x^\b)$ 
and $\l^{bc}$ obey the equation
\bea
\big{[}\x + \hf \l^{bc} M_{bc},\de_A\big{]}=0~, \qquad 
\x =\x^B \nabla_B=  \x^b \de_b + \x^{\b} \de_{\b}~,
\eea
which tells us that $\x^B$ is a Killing supervector field of $\cN=1$ AdS superspace
 \cite{KLT-M12}. This equation is equivalent to 
\bsubeq
\bea
 \de_{(\a} \x_{\b \g)}&=&0~, \qquad \de_{\b}\x^{\b \a} =- 6 \ri \x^{\a} ~, \\
\de_{\a}\x_{\b} &=& \hf \l_{\a \b}+ \cS \x_{\a \b} ~,\\
\de_{(\a} \l_{\b \g)}&=&0~, 
 \qquad \de_{\b}\l^{\b \a} =- 12 \ri \cS \x^{\a} ~. 
 \eea
\esubeq
These relations automatically follow from the (2,0) AdS Killing equations, 
eqs.~\eqref{K-eq-real11}~--~\eqref{K-eq-real14}, upon $\cN=1$ projection. 
Thus $(\x^a, \x^{\a}, \l^{ab})$ parametrise the infinitesimal isometries of 
${\rm AdS}^{3|2}$ \cite{KLT-M12} (see also\cite{KP}). 

The remaining parameters $\e^{\a}$ and $\e$ generate the second supersymmetry and ${\rm U(1)}_R$ transformations, respectively.
Using the Killing equations \eqref{K-eq-real2}, it can be shown that they satisfy the following properties 
\bsubeq \label{u1-c}
\bea
\e_{\a}= \frac{\ri}{4\cS} \de_{\a} \e ~, \qquad
\e&=& -\hf \de^{\a} \e_{\a}~,  \label{u1-c-a} \\
(\ri \de^2 + 8 \cS)\e =0~, \qquad \de_a \e &=& 0~. 
\label{u1-c-b}
\eea
\esubeq
These imply that the only independent components of $\e$ are $\e|_{\theta=0}$ and $\de_{\a}\e|_{\theta=0}$. They correspond to the  ${\rm U(1)}_R$ and
second supersymmetry transformations, respectively.

Given a matter tensor superfield $\bm U$, its (2,0) AdS transformation law
\bea
\d_\z {\bm U} = \big(\z  +\hf l^{bc} M_{bc} +\ri \t J  \big){\bm U}
\label{isometry20-real}
\eea
turns into
\bsubeq\label{isometry1-real}
\bea
\d_\z {\bm U} | &=& \d_\x {\bm U} |  +\d_\e {\bm U} | ~, \\
\d_\x {\bm U} | &=&
\Big(\x^b \de_b + \x^{\b} \de_{\b}+ \hf \l^{bc} M_{bc}\Big)
{\bm U}| ~, \\
\d_\e {\bm U} | &=& 
\Big( \e^{\b} ({\bm \de}_{\b}^{\2} \bm U)| +\ri \e J\,{\bm U}| \Big)~.
\eea
\esubeq
It follows from \eqref{2.155} and \eqref{isometry1-real}
that every $\cN=1$ descendant $U_{\a_1 \dots \a_k}:=
\big( {\bm \nabla}^\2_{{\a}_1} \cdots {\bm \nabla}^\2_{{\a}_k} {\bm U} \big)\big|$
is a tensor superfield on ${\rm AdS}^{3|2}$,
\bea
\d_\x U_{\a_1 \dots \a_k}  = \Big(\x^b \de_b + \x^{\b} \de_{\b}+ \hf \l^{bc} M_{bc}\Big)  U_{\a_1 \dots \a_k}~.
\eea
For the $\e$-transformation we get 
\bea
\d_\e U_{\a_1 \dots \a_k} &=& \e^{\b} 
\big({\bm \de}_{\b}^{\2} {\bm \nabla}^\2_{{\a}_1} \cdots {\bm \nabla}^\2_{{\a}_k} {\bm U} \big)
\big|
+ \ri \e \big( J {\bm \nabla}^\2_{{\a}_1} \cdots {\bm \nabla}^\2_{{\a}_k} {\bm U} \big)\big| \label{u1}
\\ 
&=& \e^\b U_{\b \a_1 \dots \a_k} -\e \sum_{l=1}^{k}
{\bm \nabla}^\2_{{\a}_1} \cdots 
{\bm \nabla}^\2_{\a_{l-1}}  {\bm \nabla}^\1_{\a_{l}} {\bm \nabla}^\2_{\a_{l+1}} 
\cdots {\bm \nabla}^\2_{{\a}_k} {\bm U} \big)\big|
+  \ri  q\e U_{\a_1 \dots \a_k}~,~~\non
\eea
 where $q$ is the U$(1)_R$ charge of $\bm U$ defined  by  
 $J {\bm U} =q {\bm U}$.
 In the second term on the right, we have to push ${\bm \nabla}_{\a_l}^\1$
 to the far left through the $(l-1)$ factors of ${\bm \nabla}^\2$'s
 by making use of the relation
$ \{ {\bm \nabla}_\a^\1, {\bm \nabla}_\b^\2\}= 4\ve_{\a\b}\cS J$ 
and taking into account the relation 
\bea
\big({\bm \nabla}^\1_{\a_{l}} {\bm \nabla}^\2_{{\a}_1} \cdots 
{\bm \nabla}^\2_{\a_{l-1}}  {\bm \nabla}^\2_{\a_{l+1}} 
\cdots {\bm \nabla}^\2_{{\a}_k} {\bm U} \big)\big|
= \nabla_{\a_l} U_{\a_1 \dots \a_{l-1}\a_{l+1} \dots \a_k} ~.
\eea
As the next step, the U$(1)_R$ generator $J$ should be pushed to the right until it hits
$\bm U$ producing on the way insertions of ${\bm \nabla}^\1$. Then the procedure should be repeated. As a result, the variation $ \d_\e U_{\a_1 \dots \a_k} $
is expressed in terms of the superfields $ U_{\a_1 \dots \a_{k+1}},~
 U_{\a_1 \dots \a_k} ,  \cdots U_{\a_1} , U$.

So far we have been completely general and discussed infinitely many descendants 
$U_{\a_1 \dots \a_k}$ of $\bm U$.  However only a few  of them
are functionally independent. Indeed, eq. \eqref{2_0-alg-AdS-1}
tells us  that 
\bea
&\{  {\bm \nabla}_\a^\2,  {\bm \nabla}_\b^\2\}=
2\ri  {\bm \nabla}_{\a\b}
-4\ri  \cS M_{\a\b}
~,
\eea
and thus every  $U_{\a_1 \dots \a_k}$ for $k>2$ can be expressed in terms of 
$U$, $U_\a$ and $U_{\a_1\a_2}$. Therefore, it suffices to consider $k \leq 2$.
 
Let us give two examples of matter superfields 
on (2,0) AdS superspace. We first consider a covariantly chiral scalar superfield $\bm \f,~ \bar \cD_{\a}\bm {\f} =0$, with an arbitrary U$(1)_R$ charge $q$ defined by $J \bm \f = q \bm \f$. It transforms under the (2,0) AdS isometries as
\bea
\d_{\z} \bm \f = (\z +  \ri q \t ) \bm \f~.
\eea
When expressed in the real basis \eqref{N1-deriv}, the chirality constraint on $\bm \f$ means
\bea
\bm \de^{\2}_{\a} \bm \f = \ri \bm \de^{\1}_{\a} \bm \f~,
\eea
As a result, there is only one independent ${\cal N}=1$ superfield upon reduction,
\bea
\vf := \bm \f|~.
\eea  
We then get the following relations
\bsubeq
\bea
\bm \de^{\2}_{\a} \bm \f| &=& \ri \de_{\a} \vf~, \\
( \bm \de^{\2})^2 \bm \f| &=& - \de^2 \vf - 8 \ri q \cS  \vf~.
\eea
\esubeq
The $\e$-transformation \eqref{u1} is given by 
\bea
\d_{\e} \vf = \ri \e^{\b} \de_{\b}\vf + \ri q \e  \vf~.
\eea

Our second example is a real linear superfield ${\mathbb L}= \bar{\mathbb L}~, \bar \cD^2 {\mathbb L} =0$~. The real linearity constraint relates the ${\cal N}=1$ descendants of $\mathbb{L}$ as follows:
\begin{subequations}
\bea
(\bm{\de}^{\2})^2 \mathbb{L} &=& ( \bm{\de}^{\1})^2 \mathbb{L}~,\\
\bm \de^{\1 \b} \bm \de^{\2}_{\b} \mathbb{L} &=&0~.
\label{qq1}
\eea
\end{subequations}
Thus, ${\mathbb L}$ is equivalent to two independent, real $\cN=1$ superfields:
\bea
X := {\mathbb L} |~, \qquad
W_{\a} :=\ri \bm \nabla_\a^{\2} {\mathbb L} |~.
\eea
Here $X$ is unconstrained, while $W_\a$ obeys 
the constraint \eqref{qq1}
\bea
\nabla^\a W_\a =0~,
\eea
which means that $W_\a$ is the field strength of an $\cN=1$ vector multiplet. 
Since ${\mathbb L}$ is neutral under the $R$-symmetry group U$(1)_R$, $J\, {\mathbb L} =0$, the second SUSY and U$(1)_R$ transformation laws of the ${\cal N}=1$ descendants of $\mathbb L$ are as follows:
\bsubeq
\bea
\d_{\e} X &=& \d_{\e} \mathbb{L}| = \e^{\b} (\bm \de^{\2}_{\b} \mathbb{L})| = -\ri \e^{\b} W_{\b}~,\\
\d_{\e} W_{\a} &=& \ri\, \big(\bm{ \de}^{\2}_{\a}  \d_{\e} \mathbb{L}\big)| = \ri \e^{\b}\big(\bm{ \de}^{\2}_{\b} \bm{ \de}^{\2}_{\a} \mathbb{L}\big)|- \e [J, \bm{ \de}^{\2}_{\a}] \mathbb{L}| \non\\
&=& -\e^{\b} \de_{\a \b} X - \frac{\ri}{2} \e_{\a} \de^2 X- \ri \e \de_{\a} X~.
\eea
\esubeq
 
 %%%%%%%%%%%%%%%%%%%%%%%%%%%%%%%%
 %%%%%%%%%%%%%%%%%%%%%%%%%%%%%%%%%
 
\subsection{The (2,0) AdS supersymmetric actions in $\rm AdS^{3|2}$} \label{ss2.3}

Every rigid supersymmetric field theory in (2,0) AdS superspace may be reduced to ${\cal N}=1$ AdS superspace. Here we provide the key technical details of the reduction.

In accordance with \cite{KLT-M11,BKT-M,KT-M11,KLRST-M},
there are two ways of constructing supersymmetric actions in (2,0) AdS superspace:
(i) either by integrating a real scalar 
$\cL$ over the full (2,0) AdS superspace,\footnote{The component 
inverse vierbein is defined as usual, 
$e_a{}^m (x) = E_a{}^m |_{\q=0}$, with $e^{-1}=\det(e_a{}^m)$.}
\begin{align}\label{realac}
S&= \int \rd^3x \rd^2 \q \rd^2 \bar \q \, 
{\bm E}\, \cL
= \frac{1}{16} \int \rd^3x\, e\, 
	 \cD^2 \bar \cD^2  \cL   \Big|_{\q=0}
	 = \frac{1}{16} \int \rd^3x\, e\, 
	\bar  \cD^2  \cD^2  \cL   \Big|_{\q=0}
 \\	
	&= \int \rd^3x\, e\, \Big(
	\frac{1}{16} \cD^\alpha \bar \cD^2 \cD_\alpha 
	+ \ri \cS  \bar \cD^\alpha \cD_\alpha 
	\Big)    \cL\Big|_{\q=0} 
	= \int \rd^3x\, e\, \Big(
	\frac{1}{16} \bar \cD_\alpha  \cD^2 \bar \cD^\alpha 
	+ \ri \cS  \cD^\alpha  \bar\cD_\alpha
	\Big)    \cL   \Big|_{\q=0}
	~,
\non
\end{align}
with ${\bm E}^{-1}= {\rm Ber} (E_\cA{}^\cM)$;
or (ii) by integrating a covariantly chiral scalar $\cL_{\rm c}$
over the chiral subspace of the (2,0) AdS superspace,
\begin{align}\label{chiralac}
S_{\rm c} =\int \rd^3x\, \rd^2\q\, \cE\, \cL_{\rm c}
	= -\frac{1}{4} \int \rd^3x\, e\, \cD^2 \cL_{\rm c}\Big|_{\q=0}~, \qquad
\bar\cD^\alpha \cL_{\rm c} =0~,
\end{align}
with $\cE$ being the chiral density.
The superfield Lagrangians $\cL$ and $\cL_c$ are 
neutral and charged, respectively with respect to the group ${\rm U}(1)_R$: 
\begin{align}
J \cL = 0~, \qquad J \cL_{\rm c} = -2 \cL_{\rm c}~.
\end{align}
The two types of supersymmetric actions are related to each other
by the rule
\bea
\int \rd^3x \rd^2 \q  \rd^2 \bar \q
\,{\bm E}\,\cL
= \int \rd^3x \rd^2 \q \, \cE  \cL_{\rm c}~, \qquad \cL_{\rm c}:= -\frac{1}{4} \bar {\cD}^2 \cL~.
\label{rc}
\eea

Instead of reducing the above actions to components,
in this paper we need their reduction 
to $\cN=1$ AdS superspace. 
We remind the reader that the supersymmetric action in ${\rm AdS}^{3|2}$
makes use of a real scalar Lagrangian $L$. The superspace and component forms
of the action are:
\bea
S=  \int \rd^{3|2} z \, E\,L 
=  \frac{1}{4} \int \rd^3 x \, e\,\big(\ri \de^2 +8 \cS \big) L \Big|_{\q=0}~.
\label{252}
\eea
For the action \eqref{realac} we get
\bea
S = \int \rd^3x \rd^2 \q \rd^2 \bar \q \, {\bm E}\, \cL 
= -\frac{\ri}{4}\int 
\rd^{3|2} z
\, E\, (\bm \nabla^{\2})^2 \cL \Big|~,
\label{r-action}
\eea
with $E^{-1} =\mathrm{Ber}(E_A\,^M)$. 
The chiral action \eqref{chiralac} reduces to 
${\rm AdS}^{3|2}$
as follows:
\bea
S_{\rm c} = \int \rd^3x \rd^2 \q \, \cE  \cL_{\rm c} = 2 \ri \int 
\rd^{3|2}z\, E \,  \cL_{\rm c} \Big|~.
\label{c-action}
\eea

Making use of the (2,0) AdS transformation law 
$\d \cL = \z \cL$, 
$\d \cL_{\rm c} = (\z - 2 \ri \t) \cL_{\rm c}$, and the Killing equation \eqref{Killing-real}, 
it can be checked explicitly that the ${\cN}=1$ action defined by the right-hand side of \eqref{r-action}, or \eqref{c-action} are invariant under the (2,0) AdS isometry 
transformations. 

%%%%%%%%%%%%%%%%%%%%%%%%%%
%%%%%%%%%%%%%%%%%%%%%%%%

\subsection{Supersymmetric nonlinear sigma models}

To illustrate the $(2,0)  \to (1,0)$ AdS superspace reduction described above,  
here we discuss two interesting examples.

Our first example is a general nonlinear $\s$-model with (2,0) AdS supersymmetry \cite{BKT-M,KT-M11}. It is described by the action
\bea
S = \int \rd^3x \,\rd^2 \q  \,\rd^2 \bar \q
\, {\bm E} \,K(\f^{i} , {\bar \f}^{\bar j}) + \bigg\{ \int 
\rd^3x \rd^2 \q\, \cE \,W(\f^{i})+ \rm{c.c} \bigg\}~, \qquad \bar \cD_{\a} \f^{i} =0~,
\label{2.34}
\eea
where $K(\f^i, \bar \f^{\bar j})$ is the K\"ahler potential of a K\"ahler manifold 
and $W(\f^i)$ is a superpotential. 
The U$(1)_R$ generator is realised on the dynamical superfields $\f^i $ and 
$\bar \f^{\bar i}$ as 
\bea
\ri J = {\mathfrak J}^i(\f) \pa_i + \bar{\mathfrak J}^{\bar i} (\bar \f) \pa_{\bar i} ~,
\eea
where ${\mathfrak J}^i (\f)$ is a holomorphic Killing vector field such that
\bea
{\mathfrak J}^i(\f) \pa_i K = -\frac{\ri}{2} {\mathfrak D}(\f, \bar \f) ~, \qquad 
  \bar {\mathfrak D} ={\mathfrak D}~,
   \label{2.36a}
\eea
for some  Killing potential  ${\mathfrak D}(\f, \bar \f)$. The superpotential 
has to obey the condition 
\bea
{\mathfrak J}^i(\f) \pa_i W =- 2\ri W
\eea
in order for the action \eqref{2.34} to be invariant under the (2,0) AdS isometry transformations
\bea
\d \f^i = (\z +  \ri \t J)\f^i~.
\eea

In the real representation \eqref{N1-deriv}, the chirality condition on $\f^i$ turns into
\bea
\bm \de^{\2}_{\a} \f^i = \ri \bm \de^{\1}_{\a} \f^i~.
\eea
It follows that upon ${\cal N}=1$ reduction, $\f^i$ leads to  just one superfield, 
\bea
\vf^i := \f^i|~.
\eea
In particular, we have the following relations
\bsubeq
\bea
\bm \de^{\2}_{\a} \f^i| &=& \ri \de_{\a} \vf^i~, \\
(\bm \de^{\2})^2 \f^i| &=& - \de^2 \vf^i -8 \cS {\mathfrak J}^i (\vf )~.
\eea
\esubeq
Using the reduction rules \eqref{r-action} and \eqref{c-action}, we obtain
\bea
S &=& \int \rd^{3|2} z \,E \,
\bigg\{ -\ri K_{i \bar j} (\vf, \bar \vf)
\de^{\a} \vf^{i}\de_{\a} \bar \vf^{\bar j}
 + \cS {\mathfrak D}(\vf, \bar \vf) 
+ \Big(  2\ri W(\vf) +{\rm c.c.} \Big)
\bigg\}~, ~~~
\label{N1-sigma}
\eea
where we have made use of the standard notation
\bea
K_{i_1 \cdots i_p \bar j_1 \cdots \bar j_q }:=
\frac{\pa^{p+q} K(\vf , \bar \vf)}{\pa \vf^{i_1}\cdots \pa \vf^{i_p} \pa \bar \vf^{\bar j_1} \cdots \pa \bar \vf^{\bar j_q}}~.
\eea
The action \eqref{N1-sigma} is manifestly ${\cN}=1$ supersymmetric. One may explicitly check that it is also invariant under the second supersymmetry and $R$-symmetry transformations generated by a real scalar  parameter $\e$ subject to the constraints \eqref{u1-c}, which are:
\bea
\d_{\e} \vf^i = \ri \e^{\a} \de_{\a}\vf^i 
+ \e \,{\mathfrak J}^i (\vf )~.
\eea

The family of supersymmetric $\s$-models \eqref{2.34} 
includes a special subclass which is specified by the two conditions:
(ii)   all  $\f$'s are neutral, $J \f^i=0$;  and (ii) no superpotential is present,  $W(\f)=0$. 
In this case no restriction on the K\"ahler potential is imposed by eq. \eqref{2.36a}, 
and the action \eqref{2.34} is invariant under arbitrary K\"ahler transformations 
\bea
K   \rightarrow K + \L + \bar \L, 
\eea
with  $\L(\f^i)$ a holomorphic function. The corresponding action in $\cN=1$ AdS superspace
is obtained from \eqref{N1-sigma} by setting ${\mathfrak D}(\vf, \bar \vf)=0$ and $W(\vf)=0$, 
and thus the action is manifestly K\"ahler invariant. 

Let us also consider a supersymmetric nonlinear $\s$-model formulated in terms 
of several Abelian vector multiplets with action \cite{KT-M11}
\bea
S = - 2 \int \rd^3x \,\rd^2 \q  \,\rd^2 \bar \q
\, {\bm E} \, F({\mathbb L}^i) ~, \qquad \bar \cD^2 {\mathbb L}^i =0~, \quad 
\bar{\mathbb L}^i ={\mathbb L}^i~,
\label{2.46}
\eea
where  $F(x^i)$ is a real analytic function of several variables, which is defined 
modulo linear inhomogeneous shifts
\bea
F(x) \to F(x) + b_i x^i + c~, 
\eea
with real parameters $b_i$ and $c$.
The real linear scalar ${\mathbb L}^i$ is the field strength of a vector multiplet.
Upon reduction to $\cN=1$ AdS superspace,  ${\mathbb L}^i$ 
generates two different $\cN=1$ superfields:
\bea
X^i := {\mathbb L}^i |~, \qquad
W_\a^i :=\ri \bm \nabla_\a^{\2} {\mathbb L}^i |~.
\eea
Here the real scalar $X^i$ is unconstrained, while the real spinor $W_\a^i$ obeys 
the constraint
\bea
\nabla^\a W_\a^i =0~,
\eea
which means that $W_\a^i$ is the field strength of an $\cN=1$ vector multiplet. 
Reducing the action \eqref{2.46} to $\cN=1$ AdS superspace gives
\bea
S = - \frac{\ri}{2}  \int \rd^{3|2} z\, E \, g_{ij}(X) \Big\{ 
\nabla^\a X^i \nabla_\a X^j 
+W^{\a i} W_\a^j \Big\}~,
\label{2.49}
\eea
where we have introduced the target-space metric
\bea
g_{ij} (X) = \frac{\pa^2  F(X) }{\pa X^i \pa X^j} ~.
\label{2.71}
\eea
The vector multiplets in \eqref{2.49} can be dualised into scalar ones, which gives
\bea
S_{\rm dual} = - \frac{\ri}{2}  \int \rd^{3|2} z\, E \, \Big\{ 
g_{ij} (X) 
\nabla^\a X^i \nabla_\a X^j 
+ g^{ij} (X) \nabla^\a Y_i \nabla_\a Y_j \Big\}~,
\eea
with $g^{ij} (X)$ being the inverse metric. Riemannian metrics of the type \eqref{2.71}
appeared in the literature twenty years ago
in the context of $\cN=4$  supersymmetric quantum mechanics \cite{DPRT}
and $\cN=4$ superconformal mechanics \cite{DPRST}.

%%%%%%%%%%%%%%%%%%%%%%%%%%%%%%%%%%%%
%%%%%%%%%%%%%%%%%%%%%%%%%%%%%%%%%%%%

\section{Massless higher-spin models: Type II series} \label{section3}

There exist two off-shell formulations for a massless multiplet
of half-integer superspin $(s+\hf)$ 
in (2,0) AdS superspace \cite{HK18}, with $s=2, 3, \dots, $
which are called the type II and type III series\footnote{Type I series will be referred to the longitudinal formulation 
for the gauge massless half-integer superspin multiplets in (1,1) AdS superspace
\cite{HKO} and Minkowski superspace \cite{KO}.
The type I series and its dual are naturally related to 
the off-shell formulations for massless higher-spin $\cN=1$ 
supermultiplets in four dimensions \cite{KSP,KS,KS94}.
The type II and type III series have no four-dimensional counterpart. }  
by analogy with the terminology used in \cite{KT-M11} for 
the linearised off-shell formulations for $\cN=2$ supergravity ($s=1$).
In this section we describe the $(2,0) \to (1,0)$ AdS superspace reduction 
of the type II theory. 
The reduction of the type III theory will be given in section \ref{section4}.

%%%%%%%%%%%%%%%%%%%%%%%%%%%%%%%%%%%%%%%%
%%%%%%%%%%%%%%%%%%%%%%%%%%%%%%%%%%%%%%%%%

\subsection{The type II theory}

We fix an integer $s >1$. In accordance with \cite{HK18}, the 
massless type II   multiplet of superspin $(s+\hf)$
is described in terms of two unconstrained real tensor superfields
\bea
\cV^{(\rm II)}_{(s+\hf )} = 
\Big\{ {\mathfrak H}_{\a(2s)}, \mathfrak{L}_{\a(2s-2)} \Big\} ~,
\label{3.111}
\eea
where ${\mathfrak H}_{\a(2s)} = {\mathfrak H}_{(\a_1 \dots \a_{2s})}$ 
and ${\mathfrak L}_{\a(2s-2)} = {\mathfrak L}_{(\a_1 \dots \a_{2s-2})}$
are symmetric in their spinor indices.

The dynamical superfields are defined modulo gauge transformations of the form 
\begin{subequations} \label{lambda-gauge}
\bea
\d_\l {\mathfrak H}_{\a(2s)}&=& 
{\bar \cD}_{(\a_1} \l_{\a_2 \dots \a_{2s})}-{\cD}_{(\a_1}\bar \l_{\a_2 \dots \a_{2s})}
\equiv
g_{\a(2s)}+\bar{g}_{\a(2s)} ~, 
\label{H-gauge} \\ 
\d_\l {\mathfrak L}_{\a(2s-2)} &=& -\frac{\ri}{2}
\big( \bar \cD^{\b} \l_{\b \a(2s-2)}+ \cD^{\b} \bar \l_{\b \a(2s-2)} \big)~,
\label{L-gauge}
\eea
\end{subequations}
where the gauge parameter $\l_{\a(2s-1)}$ is unconstrained complex.
Eq. \eqref{H-gauge} implies that the complex gauge parameter $g_{\a(2s)}$
is a covariantly  longitudinal linear superfield,
\bea
g_{\a(2s)}:= {\bar \cD}_{(\a_1} \l_{\a_2 \dots \a_{2s})}~, \qquad
{{\bar \cD}}_{(\a_1} g_{\a_2 \dots \a_{2s+1})} =0~. \label{ee1}
\eea
The gauge transformation of ${\mathfrak H}_{\a(2s)}$, eq. \eqref{H-gauge},
corresponds to the superconformal  gauge prepotential \cite{KO,HKO}. 
The  prepotential ${\mathfrak L}_{\a(2s-2)} $ is a compensating multiplet.
In addition to \eqref{L-gauge}, the compensator ${\mathfrak L}_{\a(2s-2)}$ also possesses its own gauge freedom of the form 
\bea
\d_\x {\mathfrak L}_{\a(2s-2)} 
=  { \x}_{\a(2s-2)}+ \bar { \x}_{\a(2s-2)} ~, \qquad  \bar \cD_{\b} \x_{\a(2s-2)}=0~,
\label{prep-gauge}
\eea
with the gauge parameter ${\x_{\a(2s-2)}}$ being covariantly chiral.

Associated with ${\mathfrak L}_{\a(2s-2)} $
is the real field strength 
\bea
 \mathbb{L}_{\a(2s-2)} = \ri \cD^{\b} \bar \cD_{\b} {\mathfrak L}_{\a(2s-2)} ~,
 \qquad \mathbb{L}_{\a(2s-2)}= \bar{\mathbb{L}}_{\a(2s-2)}~,
\label{2.2}
\eea
which is a covariantly linear superfield, 
\bea
{\cD}^2 \mathbb{L}_{\a(2s-2)}=0 \quad \Longleftrightarrow \quad 
\bar \cD^2 \mathbb{L}_{\a(2s-2)}=0~.
\label{RLconst}
\eea
It is inert under the gauge transformation \eqref{prep-gauge}, $\d_\x  \mathbb{L}_{\a(2s-2)} =0$. From \eqref{L-gauge}
we can read off the $\l$-gauge transformation of the field strength: 
\bea
\d_\l \mathbb{L}_{\a(2s-2)}&=&
\frac{1}{4}
\big( \cD^{\b} {\bar \cD}^2 \l_{\b \a(2s-2)}- \bar \cD^{\b} {\cD}^2 \bar \l_{\b \a(2s-2)}\big) ~.~~~
\non \\
\qquad &=& -\frac{s}{2s+1} \cD^{\b} {\bar \cD}^{\g}\big(g_{\b \g \a(2s-2)} + \bar g_{\b \g \a(2s-2)}\big)
-\frac{2 \ri s}{2s+1} \cD^{\b \g} \bar g_{\b \g \a(2s-2)} ~.
\label{bbL-gauge}
\eea

The type II theory is described by the action
\bea
S^{(\rm II)}_{(s+\hf)} [{\mathfrak H}_{\a(2s)} ,{\mathfrak L}_{\a(2s-2)} ]
&=& \Big(-\hf \Big)^{s} 
\int 
\rd^3x \rd^2 \q  \rd^2 \bar \q
\, {\bm E}\, \bigg\{\frac{1}{8}{\mathfrak H}^{\a(2s)}
\cD^{\b}\bar{\cD}^{2} \cD_{\b}
{\mathfrak H}_{\a(2s)} \non \\
&&-\frac{s}{8}([\cD_{\b},\bar{\cD}_{\g}]{\mathfrak H}^{\b \g \a(2s-2)})
[\cD^{\d},\bar{\cD}^{\r}]{\mathfrak H}_{\d \r \a(2s-2)}
 \non \\
&& +\frac{s}{2}(\cD_{\b \g}{\mathfrak H}^{\b \g \a(2s-2)})
\cD^{\d \r}{\mathfrak H}_{\d \r \a(2s-2)}+ 2\ri s \,{\cS} {\mathfrak H}^{\a(2s)} {\cD}^\b {\bar \cD}_{\b} {\mathfrak H}_{\a(2s)}
\non \\
&&-  \frac{2s-1}{2} \Big( \mathbb{L}^{\a(2s-2)} [\cD^{\b}, \bar \cD^{\g}] {\mathfrak H}_{\b \g \a(2s-2)}
+ 2
\mathbb{L}^{\a(2s-2)} \mathbb{L}_{\a(2s-2)} \Big) \non\\
&&-\frac{(s-1)(2s-1)}{4s} \Big(\cD_{\b} \mathfrak{L}^{\b \a(2s-3)} \bar \cD^2 \cD^{\g} \mathfrak{L}_{\g \a(2s-3)}+\mathrm{c.c.} \Big)\non\\
&& -4(2s-1) \cS \mathfrak{L}^{\a(2s-2)} \mathbb{L}_{\a(2s-2)}
\bigg\}~.
\label{action}
\eea
It is invariant under the gauge transformations \eqref{lambda-gauge} and \eqref{prep-gauge}.

The structure $\cD_{\b} \mathfrak{L}^{\b \a(2s-3)}
 \bar \cD^2 \cD^{\g} \mathfrak{L}_{\g \a(2s-3)}$ in \eqref{action} is not defined  for $s=1$. 
However it comes with the factor $(s-1)$ and therefore 
drops out from \eqref{action}
for $s=1$. The action \eqref{action} for $s=1$ coincides with the linearised 
action for (2,0) AdS supergravity, which was originally derived 
in section 10.1 of \cite{KT-M11}.

%%%%%%%%%%%%%%%%%%%%%%%%%%%%%%%%%%%%
%%%%%%%%%%%%%%%%%%%%%%%%%%%%%%%%%%%%%

\subsection{Reduction of the gauge prepotentials to ${\rm AdS}^{3|2}$}

Let us turn to reducing the gauge prepotentials \eqref{3.111} 
to $\cN=1$ AdS superspace.\footnote{In the super-Poincar\'e case,
the $\cN=2 \to \cN=1$ reduction of ${\mathfrak H}_{\a(2s)}$ 
 has been carried out in \cite{KT}.} 
 Our first task is to work out such a reduction for 
 the superconformal gauge multiplet ${\mathfrak H}_{\a(2s)}$.
In the real representation \eqref{N1-deriv}, the longitudinal linear constraint \eqref{ee1} takes the form
\be 
{\bm \nabla}^{\underline{2}}_{(\a_1} g_{\a_2 \dots \a_{2s+1})} 
= \ri {\bm \nabla}^{\underline{1}} _{(\a_1}g_{\a_2 \dots \a_{2s+1})}~.
\ee
It follows that $g_{\a(2s)}$ has 
two independent $\theta_{\underline{2}}$-components, which are
\bea
g_{\a(2s)} |~, \qquad {\bm \nabla}^{\underline{2}\, \b}g_{\a(2s-1)\b}|~.
\eea
The gauge transformation of ${\mathfrak H}_{\a(2s)}$, eq. \eqref{H-gauge}, allows us to choose two gauge conditions
\bea
{\mathfrak H}_{\a(2s)}| =0~, \qquad \
{\bm \nabla}^{\underline{2}\, \b}{\mathfrak H}_{\a(2s-1)\b}| =0~.
\label{q1}
\eea
In this gauge we stay with the following unconstrained real ${\cN=1}$ superfields:
\begin{subequations} \label{q2}
\bea
H_{\a(2s+1)} &:=& {\ri} {\bm \nabla}^{\underline{2}}_{(\a_1} {\mathfrak H}_{\a_2 \dots \a_{2s+1})} | ~, \label{H1} \\
H_{\a(2s)} &:=& \frac{\ri}{4}({\bm \nabla}^{\underline{2}})^2 {\mathfrak H}_{\a(2s)}|~. \label{H2}
\eea
\end{subequations}
There exists a residual gauge freedom which preserves the gauge conditions \eqref{q1}. It is described by unconstrained real ${\cN=1}$ superfields $\z_{\a(2s)}$ and $\z_{\a(2s-1)}$ defined by 
\begin{subequations}
\bea
g_{\a(2s)} | &=& -\frac{\ri}{2} \z_{\a(2s)}~,\,\,  \qquad \qquad {\bar \z}_{\a(2s)} = \z_{\a(2s)}~, \label{q3.a} \\
\bm \nabla^{\underline{2}\, \b} g_{\a(2s-1) \b} | &=& \frac{2s+1}{2s} \z_{\a(2s-1)} ~, \qquad {\bar \z}_{\a(2s-1)} = \z_{\a(2s-1)}~. \label{q3.b}
\eea
\end{subequations}
The gauge transformation laws of the superfields \eqref{q2} are given by
\begin{subequations} \label{q4.a-b}
\bea
\d H_{\a(2s+1)} &=& {\ri} \nabla_{(\a_1} \z_{\a_2 \dots \a_{2s+1})}~, \label{q4.a}\\
\d H_{\a(2s)} &=&  \nabla_{(\a_1}\z_{\a_2 \dots \a_{2s})} \label{q4.b}~.
\eea
\end{subequations}

Our next step is  to reduce 
the compensator ${\mathfrak L}_{\a(2s-2)}$ to ${\cN=1}$ AdS superspace. 
Making use of the representation \eqref{N1-deriv}, we observe that the chirality condition \eqref{prep-gauge} reads
\bea
{\bm \nabla}^{\underline{2}}_{\b} \x_{\a(2s-2)}= \ri {\bm \nabla}^{\underline{1}}_{\b} \x_{\a(2s-2)}~. \label{l1}
\eea
The gauge transformation \eqref{prep-gauge} allows us to impose a gauge condition
\bea
{\mathfrak L}_{\a(2s-2)}| =0~. \label{l2}
\eea
Thus, upon reduction to ${\cal N}=1$ superspace, we have the following real superfields
\begin{subequations}
\bea
\J_{\b; \, \a(2s-2)} &:=& \ri {\bm \nabla}^{\underline{2}}_{\b} {\mathfrak L}_{\a(2s-2)}|~, \label{l3} \\
L_{\a(2s-2)} &:=& \frac{\ri}{4}({\bm \nabla}^{\underline{2}})^2 {\mathfrak L}_{\a(2s-2)}|~. \label{l4}
\eea
\end{subequations}
Here $\J_{\b; \, \a(2s-2)}$ is a reducible superfield which belongs
to the representation 
 $ {\bf 2}\otimes (\bf{2s- 1}) $ of ${\rm SL} (2,{\mathbb R})$,  
$\J_{\b; \, \a_1 \dots \a_{2s-2}} = \J_{\b; \, (\a_1 \dots \a_{2s-2})}$. 
The condition \eqref{l2} is preserved by the residual gauge freedom generated by a real unconstrained ${\cal N}=1$ superfield $\eta_{\a(2s-2)}$ defined by
\bea
\x_{\a(2s-2)}| = -\frac{\ri}{2} \eta_{\a(2s-2)}~, \qquad \bar \eta_{\a(2s-2)} = \eta_{\a(2s-2)}~. \label{l5}
\eea
We may now determine how the $\eta$-transformation acts on the superfields 
\eqref{l3} and \eqref{l4}. We obtain
\begin{subequations}
\bea
\d_{\eta} \J_{\b; \, \a(2s-2)} &=& \ri \nabla_{\b} \eta_{\a(2s-2)}~, \label{l6} \\
\d_{\eta} L_{\a(2s-2)} &=& 0~,
\eea
\end{subequations}
where we have used the chirality constraint \eqref{l1} and the expression \eqref{l5} for the residual gauge transformation. 

Next, we analyse the $\l$-gauge transformation and reduce the ${\cal N}=2$ field strength $\mathbb{L}_{\a(2s-2)}$ to ${\rm AdS}^{3|2}$. In the real basis for the covariant derivatives, the real linearity constraint \eqref{RLconst} is equivalent to two constraints:
\begin{subequations}
\bea
({\bm \nabla}^{\underline{2}})^2 \mathbb{L}_{\a(2s-2)} 
&=& ({\bm \nabla}^{\underline{1}})^2 \mathbb{L}_{\a(2s-2)}~, \\
{\bm \nabla}^{\underline{1}\, \b} {\bm \nabla}^{\underline{2}}_{\b} \mathbb{L}_{\a(2s-2)} &=& 0~. \label{BI}
\eea
\end{subequations}
These constraints imply that the resulting ${\cal N}=1$ components of $\mathbb{L}_{\a(2s-2)}$ are given by
\bea
\mathbb{L}_{\a(2s-2)}|~, \qquad \quad \ri 
{\bm \nabla}^{\underline{2}}_{\b} \mathbb{L}_{\a(2s-2)}|~,
\eea
of which the former is unconstrained and the latter is a constrained ${\cN=1}$ 
superfield that proves to be  a gauge-invariant field strength, as we shall see below.
The relation between $\mathbb{L}_{\a(2s-2)}$ and the prepotential ${\mathfrak L}_{\a(2s-2)}$ is given by \eqref{2.2}, which can be expressed as
\bea
\mathbb{L}_{\a(2s-2)} = -\frac{\ri}{2}\Big\{ ({\bm \nabla}^{\underline{1}})^2 
+ ( {\bm \nabla}^{\underline{2}})^2\Big\} {\mathfrak L}_{\a(2s-2)}~. 
\label{bbl1}
\eea
We now compute the bar-projection of \eqref{bbl1} in the gauge \eqref{l2} and make use of the definition \eqref{l4} to obtain
\bea
\mathbb{L}_{\a(2s-2)}| = -2 L_{\a(2s-2)}~.
\eea
Making use of \eqref{bbl1} and \eqref{l3}, the bar-projection of 
$\ri {\bm \nabla}^{\underline{2}}_{\b} \mathbb{L}_{\a(2s-2)}$ leads to the ${\cal N}=1$ field strength
\bea
 \cW_{\b;\, \a(2s-2)}&:=& \ri {\bm \nabla}^{\underline{2}}_{\b} \mathbb{L}_{\a(2s-2)}|
 =
  -\ri \Big( \nabla^{\g} \nabla_{\b} - 4\ri \cS \delta^{\g}_{\,\b}\Big){\Psi}_{\g; \,\a(2s-2)} ~.
 \label{eqW}
\eea
Here $\cW_{\b;\, \a(2s-2)}$ is a real superfield, $\cW_{\b;\, \a(2s-2)}= {\bar \cW}_{\b;\, \a(2s-2)}$, and is a descendant of the real unconstrained prepotential $\J_{\b; \, \a(2s-2)}$ defined modulo gauge transformation \eqref{l6}. The field strength proves to be gauge invariant under \eqref{l6},  and it satisfies the condition 
\bea
\nabla^{\b}\cW_{\b;\, \a(2s-2)}=0~,
\eea
as a consequence of \eqref{BI} and the identity \eqref{A2}. 
Let us express the gauge transformation of $\mathbb{L}_{\a(2s-2)}$, eq.~\eqref{bbL-gauge} in terms of the real basis for the covariant derivatives. This leads to 
\bea
\d  \mathbb{L}_{\a(2s-2)} &=& \frac{\ri s}{2s+1} \Big\{ {\bm \nabla}^{\1 \b} {\bm \nabla}^{\underline{2}\g}\Big( g_{\b \g \a(2s-2)} + {\bar g}_{\b \g \a(2s-2)}\Big)
\non \\
&+& {\bm \nabla}^{\b \g}\Big( g_{\b \g \a(2s-2)} - {\bar g}_{\b \g \a(2s-2)}\Big) \Big\}~, 
\eea
In a similar way, one should also rewrite 
${\bm \nabla}^{\2}_{\b} \,\d \mathbb{L}_{\a(2s-2)}$ in the real basis.
This allows us to derive the gauge transformations 
 for $L_{\a(2s-2)}$ and $\cW_{\b;\, \a(2s-2)}$
\begin{subequations}
\bea
\d L_{\a(2s-2)} &=& -\frac{s}{2(2s+1)} \nabla^{\b \g} \z_{\b \g \a(2s-2)}~, \label{N1-Lgauge} \\
\d \cW_{\b; \, \a(2s-2)} &=& \ri \big( \nabla^{\g} \nabla_{\b} -4 \ri \cS \d^{\g}_{\,\b} \big) \z_{\g \a(2s-2)}~. \label{eqWgauge}
\eea
We can then read off the transformation law for the prepotential $\J_{\b; \, \a(2s-2)}$
\bea
\d \J_{\b; \,\a(2s-2)} &=& -\z_{\b \a(2s-2)}+ \ri \nabla_{\b} \eta_{\a(2s-2)}~, 
\label{N1-psigauge}
\eea
\end{subequations}
where we have also taken into account the $\eta$-gauge freedom \eqref{l6}.

Applying the ${\cal N}=1$ reduction rule \eqref{r-action} to the type II action \eqref{action}, we find that it becomes a sum of two actions,
\bea
S^{(\rm II)}_{(s+\hf)}[{\mathfrak H}_{\a(2s)} ,{\mathfrak L}_{\a(2s-2)} ] 
= S^{\parallel}_{(s+\hf)}[H_{\a(2s+1)} ,L_{\a(2s-2)} ] 
+ S^{\perp}_{(s)}[H_{\a(2s)} ,{\Psi}_{\b; \,\a(2s-2)} ]~.
\eea
Explicit expressions for these ${\cal N}=1$ actions will be given in the next subsection. 

%%%%%%%%%%%%%%%%%%%%%%%%%%%%%%%%%%%%%
%%%%%%%%%%%%%%%%%%%%%%%%%%%%%%%%%%%%%%%%

\subsection{Massless higher-spin $\cN=1$ supermultiplets in AdS${}_3$} \label{ss32}

The gauge transformations 
 \eqref{q4.a}, \eqref{q4.b},
  \eqref{N1-Lgauge} and \eqref{N1-psigauge} tell us that in fact 
  we are dealing with two different $\cN=1$ supersymmetric 
  higher-spin gauge theories. 
  
   Given a positive integer $n>0$,
  we say that a supersymmetric gauge theory describes a 
  multiplet of superspin $n/2$ if it is formulated in terms of 
  a superconformal gauge prepotential $H_{\a(n)}$ and possibly a compensating 
  multiplet.  The gauge freedom of the real tensor superfield $H_{\a(n)}$ is 
  \bea 
 \d_\z H_{\a(n)}  = \ri^n (-1)^{ \left \lfloor{n/2}\right \rfloor }
 \nabla_{(\a_1} \z_{\a_2 \dots \a_n)}~,
 \label{3.299}
 \eea
 with the gauge parameter $\z_{\a(n-1)}$ being real but otherwise 
 unconstrained. 
  
 %%%%%%%%%%%%%%%%%%%%%%%%%%%%%%%%%%
 %%%%%%%%%%%%%%%%%%%%%%%%%%%%%%%%%%% 
  
\subsubsection{Longitudinal
formulation for massless superspin-$(s+\hf)$ multiplet}

  One of the two $\cN=1$ theories  provides an off-shell formulation 
  for the massless superspin-$(s+\hf)$ multiplet. It is formulated in terms of the real unconstrained gauge superfields
\bea
\cV^{\parallel}_{(s+\hf )} = \Big\{H_{\a(2s+1)},\, L_{\a(2s-2)} \Big\} ~, 
\label{hf}
\eea
which are defined modulo gauge transformations 
\begin{subequations} \label{hf-gauge}
\bea
\d H_{\a(2s+1)} &=& \ri \nabla_{(\a_1} \z_{\a_2 \dots \a_{2s+1})}~,\\
\d L_{\a(2s-2)} &=& -\frac{s}{2(2s+1)} \nabla^{\b \g} \z_{\b \g \a(2s-2)}~,
\eea
\end{subequations}
where the parameter $\z_{\a(2s)}$ is unconstrained real.
The  gauge-invariant  action is
\bea
\lefteqn{S^{\parallel}_{(s+\hf)}[H_{\a(2s+1)} ,L_{\a(2s-2)} ]
= \Big(-\hf \Big)^{s} 
\int \rd^{3|2} z
\, E\, \bigg\{-\frac{\ri}{2} H^{\a(2s+1)} {\mathbb{Q}} H_{\a(2s+1)} }
\non \\
&&\qquad -\frac{\ri}{8} \nabla_{\b} H^{\b \a(2s)} \nabla^2 \nabla^{\g}H_{\g \a(2s)}+\frac{\ri s}{4}{\nabla}_{\b \g}H^{\b \g \a(2s-1)} {\nabla}^{\rho \d}H_{\rho \d \a(2s-1)}
\non \\
&& \qquad + (2s-1) L^{\a(2s-2)}\nabla^{\b \g} \nabla^{\d} H_{\b \g \d \a(2s-2)}
\non \\
&& \qquad + 2 (2s-1)\Big( L^{\a(2s-2)} (\ri \nabla^2 - 4 \cS) L_{\a(2s-2)}
- \frac{\ri}{s}(s-1) \nabla_{\b} L^{\b \a(2s-3)} \nabla^{\g}L_{\g \a(2s-3)}\Big)
\non \\
&& \qquad + \cS \Big(s \,\de_{\b}H^{\b \a(2s)} \de^{\g} H_{\g \a(2s)}+ \hf (2s+1)H^{\a(2s+1)}( \nabla^2-4 \ri \cS)H_{\a(2s+1)} \Big)
\bigg\}~,
\label{action-t2-half}
\eea
where ${\mathbb{Q}}$ is the quadratic Casimir operator of the 3D ${\cal N}=1$ AdS supergroup \eqref{casimir}~.
The action \eqref{action-t2-half} coincides with the off-shell ${\cal N}=1$ supersymmetric action for massless half-integer superspin in AdS in the form given in \cite{KP}.
This supersymmetric gauge theory in ${\rm AdS}^{3|2}$ was described in 
\cite{KP}. Its flat-superspace limit was presented earlier in \cite{KT}.
In what follows, we will refer to the above theory as the longitudinal
formulation for the massless superspin-$(s+\hf)$ multiplet. 

The structure $\nabla_{\b} L^{\b \a(2s-3)} \nabla^{\g}L_{\g \a(2s-3)}$ in 
\eqref{action-t2-half} is not defined for $s=1$. However it comes with the factor 
$(s-1)$ and drops out from \eqref{action-t2-half} for $s=1$. The resulting action 
\bea
S^{\parallel}_{(\frac 32)} [H_{\a(3)} ,L ]
&=& -\hf \int \rd^{3|2} z
\, E\, \bigg\{-\frac{\ri}{2} H^{\a(3)} {\mathbb{Q}} H_{\a(3)} 
-\frac{\ri}{8} \nabla_{\b} H^{\b \a(2)} \nabla^2 \nabla^{\g}H_{\g \a(2)}
\non \\
&&
+\frac{\ri }{4}{\nabla}_{\b \g}H^{\b \g \a} {\nabla}^{\rho \d}H_{\rho \d \a}
+  L \nabla^{\b \g} \nabla^{\d} H_{\b \g \d }
 + 2  L \big( \ri \nabla^2  -4\cS\big) L
 \\
&& 
+ \cS \Big( \de_{\b}H^{\b \a(2)} \de^{\g} H_{\g \a(2)}
+ \frac 32  H^{\a(3)}\big( \nabla^2 -4\ri \cS\big) H_{\a(3)} 
 \Big) \bigg\}
\non
\eea
is the linearised action for $\cN=1$ AdS supergravity. 
In the flat-superspace limit, the action is equivalent to the one given in \cite{GGRS}.

%%%%%%%%%%%%%%%%%%%%%%%%%%%%%%%%%%
 %%%%%%%%%%%%%%%%%%%%%%%%%%%%%%%%%%% 
  
\subsubsection{Transverse formulation for massless superspin-$s$ multiplet}

The other $\cN=1$ theory provides a formulation 
  for the massless superspin-$s$ multiplet. 
 It is described by the unconstrained real  superfields 
\bea
\cV^{\perp}_{(s)} = \Big\{H_{\a(2s)}, \J_{\b; \, \a(2s-2)} \Big\} ~,
\label{t2new}
\eea
which are defined modulo gauge transformations of the form
\begin{subequations} \label{t2new-gauge}
\bea
\d H_{\a(2s)} &=& \nabla_{(\a_1} \z_{\a_2 \dots \a_{2s})}~,\\
\d \J_{\b; \,\a(2s-2)} &=& -\z_{\b \a(2s-2)}+ \ri \nabla_{\b} \eta_{\a(2s-2)}~,
\eea
\end{subequations}
where the gauge parameters $\z_{\a(2s-1)}$ and $\eta_{\a(2s-2)}$ 
are unconstrained real.
The gauge-invariant action is given by
\begin{subequations} \label{action-t2-new}
\bea
\lefteqn{S^{\perp}_{(s)}[H_{\a(2s)} ,{\Psi}_{\b; \,\a(2s-2)} ]
= \Big(-\hf \Big)^{s} 
\int \rd^{3|2}z\,
 E\, \bigg\{\frac{1}{2} H^{\a(2s)} (\ri \nabla^2 +8 s \cS) H_{\a(2s)}}
\non \\
&& \qquad \qquad - \ri s \nabla_{\b} H^{\b \a(2s-1)} \nabla^{\g}H_{\g \a(2s-1)} 
-(2s-1) \cW^{\b ;\,\a(2s-2)} \nabla^{\g} H_{\g \b \a(2s-2)}
\non \\
&& \qquad \qquad -\frac{\ri}{2} (2s-1)\Big(\cW^{\b ;\, \a(2s-2)} \cW_{\b ;\, \a(2s-2)}+\frac{s-1}{s} \cW_{\b;}\,^{\b \a(2s-3)} \cW^{\g ;}\,_{\g \a(2s-3)} \Big) 
\non\\
&& \qquad \qquad
 -2 \ri (2s-1) \cS \Psi^{\b ;\, \a(2s-2)} \cW_{\b ; \, \a(2s-2)}
\bigg\}~,
\label{action-t2-new-a}
\eea
where $\cW_{\b; \, \a(2s-2)}$ denotes the field strength
\bea
 \cW_{\b;\, \a(2s-2)} =
  -\ri \Big( \nabla^{\g} \nabla_{\b} - 4\ri \cS \delta^{\g}_{\,\b}\Big){\Psi}_{\g; \,\a(2s-2)} ~,
  \qquad \nabla^\b  \cW_{\b;\, \a(2s-2)}=0~.
\eea
\end{subequations}
The action \eqref{action-t2-new} defines a new ${\cal N}=1$ supersymmetric higher-spin theory which did not appear in \cite{KP,HKO, KT}
even in the super-Poincar\'e case.

The structure $\cW_{\b;}\,^{\b \a(2s-3)} \cW^{\g ;}\,_{\g \a(2s-3)} $
in \eqref{action-t2-new-a} is not defined for $s=1$. 
However it comes with the factor $(s-1)$ and drops out from 
\eqref{action-t2-new-a} for $s=1$.
The resulting gauge-invariant action 
\bea
S^{\perp}_{(1)}[H_{\a(2)} ,{\Psi}_{\b} ]
&=& -\hf \int \rd^{3|2}z\, E\, \bigg\{\frac{1}{2} H^{\a(2)} (\ri \nabla^2 + 8 \cS) H_{\a(2)}
- \ri  \nabla_{\b} H^{\b \a } \nabla^{\g}H_{\g \a} 
\non \\
&& 
\qquad - \cW^{\b} \nabla^{\g} H_{\g \b } -\frac{\ri}{2} \cW^{\b } \cW_{\b}
 -2 \ri  \cS \Psi^{\b } \cW_{\b }
\bigg\}
\label{337}
\eea
provides an off-shell realisation for a massless gravitino multiplet in AdS${}_3$.
In the flat-superspace limit, this model reduces to the one described in 
\cite{KT}.

In the $s>1$ case, the gauge freedom of the prepotential $\J_{\b ;\, \a(2s-2)}$ \eqref{t2new-gauge} allows us to impose a gauge condition 
\bea
\J_{(\a_1;\, \a_2 \dots \a_{2s-1})} =0 \quad \Longleftrightarrow \quad 
\J_{\b ;\, \a(2s-2)} =  \sum_{k=1}^{2s-2}\ve_{\b \a_k} \vf_{\a_1 \dots \hat{\a}_k \dots \a_{2s-2}}~,
\label{3.355}
\eea
for some field  $\vf_{\a(2s-3)}$.
Since we gauge away the symmetric part of $\J_{\b ;\, \a(2s-2)}$, the two gauge parameters $\z_{\a(2s-1)}$ and $\eta_{\a(2s-2)}$ are related. The theory is now realised in terms of the following dynamical variables
\bea
\Big\{H_{\a(2s)},~\vf_{\a(2s-3)}\Big\}~,
\eea
with the gauge freedom
\begin{subequations}
\bea
\d H_{\a(2s)} &=& -\nabla_{(\a_1 \a_2} \eta_{\a_3 \dots \a_{2s})}~,\\
\d \vf_{\a(2s-3)}&=& \ri \nabla^{\b} \eta_{\b \a(2s-3)}~.
\eea
\end{subequations}
It follows that in the flat-superspace limit, $\cS=0$, 
and in the gauge \eqref{3.355},
the action \eqref{action-t2-new} coincides with eq.~(B.25) of \cite{HKO}. The component structure of this model will be discussed in Appendix B.1.

%%%%%%%%%%%%%%%%%%%%%%%%%%%%%%%%%%%%%%%
%%%%%%%%%%%%%%%%%%%%%%%%%%%%%%%%%%%%%%

\section{Massless higher-spin models: Type III series} \label{section4}

In this section we carry out  the ${\cN=1}$ AdS superspace reduction of the type III theory \cite{HK18} following the procedure  employed in section \ref{section3}. 

%%%%%%%%%%%%%%%%%%%%%%%%%%%%%%

\subsection{The type III theory}

We fix a positive integer $s>1$. 
In accordance with \cite{HK18}, the 
massless type III multiplet of  superspin $(s+\hf)$ 
is described in terms of two unconstrained real tensor superfields
\bea
\cV^{(\rm III)}_{(s+\hf )} = 
\Big\{ {\mathfrak H}_{\a(2s)}, \mathfrak{V}_{\a(2s-2)} \Big\} ~,
\eea
which are symmetric in their spinor indices, 
${\mathfrak H}_{\a(2s)} = {\mathfrak H}_{(\a_1 \dots \a_{2s})}$ 
and ${\mathfrak V}_{\a(2s-2)} = {\mathfrak V}_{(\a_1 \dots \a_{2s-2})}$.

The dynamical superfields are defined modulo gauge transformations of the form 
\begin{subequations} \label{3.1}
\bea
\d_\l {\mathfrak H}_{\a(2s)}&=& 
{\bar \cD}_{(\a_1} \l_{\a_2 \dots \a_{2s})}-{\cD}_{(\a_1}\bar \l_{\a_2 \dots \a_{2s})}
=g_{\a(2s)}+\bar{g}_{\a(2s)} ~, 
\label{3.1a} \\ 
\d_\l {\mathfrak V}_{\a(2s-2)} &=& \frac{1}{2s}
\big( \bar \cD^{\b} \l_{\b \a(2s-2)} - \cD^{\b} \bar \l_{\b \a(2s-2)} \big)~,
\label{3.1b}
\eea
\end{subequations}
where the gauge parameter $\l_{\a(2s-1)}$ is unconstrained complex, 
and the longitudinal linear parameter $g_{\a(2s)}$ 
is defined as in \eqref{ee1}.
As in the type II case, ${\mathfrak H}_{\a(2s)}$ is the superconformal 
gauge multiplet, while ${\mathfrak V}_{\a(2s-2)} $ is a compensating multiplet.
The only difference from the type II case occurs in the gauge transformation law for the compensator ${\mathfrak V}_{\a(2s-2)}$.

The compensator ${\mathfrak V}_{\a(2s-2)}$ also possesses its own gauge freedom of the form 
\bea
\d_\x {\mathfrak V}_{\a(2s-2)} 
=  { \x}_{\a(2s-2)}+ \bar { \x}_{\a(2s-2)} ~, \qquad  \bar \cD_{\b} \x_{\a(2s-2)}=0~,
\label{3.3}
\eea
with the gauge parameter ${\x_{\a(2s-2)}}$ being covariantly chiral, 
but otherwise arbitrary. 

Associated with ${\mathfrak V}_{\a(2s-2)} $
is the real field strength 
\bea
 \mathbb{V}_{\a(2s-2)} = \ri \cD^{\b} \bar \cD_{\b} {\mathfrak V}_{\a(2s-2)} ~,
 \qquad \mathbb{V}_{\a(2s-2)}= \bar{\mathbb{V}}_{\a(2s-2)}~,
\label{3.4}
\eea
which is inert under \eqref{3.3}, $\d_\x  \mathbb{V}_{\a(2s-2)} =0$. It is not difficult to see that 
$\mathbb{V}_{\a(2s-2)}$ is covariantly linear, 
\bea
{\cD}^2 \mathbb{V}_{\a(2s-2)}=0 \qquad \Longleftrightarrow \quad  \bar {\cD}^2 \mathbb{V}_{\a(2s-2)}=0~. \label{3.5}
\eea
It varies under the $\l$-gauge transformation as 
\bea
\d_\l \mathbb{V}_{\a(2s-2)}&=&
\frac{\ri}{4s}
\big( \cD^{\b} {\bar \cD}^2 \l_{\b \a(2s-2)}+ \bar \cD^{\b} {\cD}^2 \bar \l_{\b \a(2s-2)}\big) ~.~~~
\non \\
\qquad &=& -\frac{\ri}{2s+1} \cD^{\b} {\bar \cD}^{\g}\big(g_{\b \g \a(2s-2)} - \bar g_{\b \g \a(2s-2)}\big)
-\frac{2}{2s+1} \cD^{\b \g} \bar g_{\b \g \a(2s-2)} ~.
\label{3.6}
\eea

Modulo normalisation, there exists a unique action being
invariant under the gauge transformations \eqref{3.1} and \eqref{3.3}.
 It is given by
\bea
S^{(\rm III)}_{(s+\hf)}  [{\mathfrak H}_{\a(2s)} ,{\mathfrak V}_{\a(2s-2)} ]
&=& \Big(-\hf \Big)^{s} 
\int 
\rd^3x \rd^2 \q  \rd^2 \bar \q
\, {\bm E}\, \bigg\{\frac{1}{8}{\mathfrak H}^{\a(2s)}
\cD^{\b}\bar{\cD}^{2} \cD_{\b}
{\mathfrak H}_{\a(2s)} \non \\
&&-\frac{1}{16}([\cD_{\b},\bar{\cD}_{\g}]{\mathfrak H}^{\b \g \a(2s-2)})
[\cD^{\d},\bar{\cD}^{\r}]{\mathfrak H}_{\d \r \a(2s-2)}
 \non \\
&& +\frac{1}{4}(\cD_{\b \g}{\mathfrak H}^{\b \g \a(2s-2)})
\cD^{\d \r}{\mathfrak H}_{\d \r \a(2s-2)}+ \ri  \,{\cS} {\mathfrak H}^{\a(2s)} {\cD}^\b {\bar \cD}_{\b} {\mathfrak H}_{\a(2s)}
\non \\
&&-  \frac{2s-1}{2} \Big( \mathbb{V}^{\a(2s-2)} \cD^{\b \g} {\mathfrak H}_{\b \g \a(2s-2)}
+ \frac{1}{2}
\mathbb{V}^{\a(2s-2)} \mathbb{V}_{\a(2s-2)} \Big) \non\\
&&+\frac{1}{8}(s-1)(2s-1)\Big(
\cD_{\b} \mathfrak{V}^{\b \a(2s-3)} \bar \cD^2 \cD^{\g} \mathfrak{V}_{\g \a(2s-3)}+ \mathrm{c.c.}\Big) \non\\
&&+2s(2s-1) \cS \mathfrak{V}^{\a(2s-2)} \mathbb{V}_{\a(2s-2)}
\bigg\}~.
\label{action2}
\eea
Although the structure 
$\cD_{\b} \mathfrak{V}^{\b \a(2s-3)} \bar \cD^2 \cD^{\g} \mathfrak{V}_{\g \a(2s-3)}$
in \eqref{action2} is not defined for $s=1$, it comes with the factor $(s-1)$ 
and drops out from \eqref{action2} for the $s=1$ case.
In this case the action 
coincides with the type III supergravity 
action in (2,0) AdS superspace, which was originally derived 
in section 10.2 of \cite{KT-M11}.

%%%%%%%%%%%%%%%%%%%%%%%%%%%%%%%%%%%
%%%%%%%%%%%%%%%%%%%%%%%%%%%%%%%%%%%

\subsection{Reduction of the gauge prepotentials to ${\rm AdS}^{3|2}$}

The reduction of the superconformal gauge multiplet ${\mathfrak H}_{\a(2s)}$ to  ${\rm AdS}^{3|2}$ has been carried out in the previous section. We saw that in the gauge \eqref{q1}, ${\mathfrak H}_{\a(2s)}$ is described by the two unconstrained real superfields $H_{\a(2s+1)}$ and $H_{\a(2s)}$ defined according to \eqref{q2}, with their gauge transformation laws given by eqs. \eqref{q4.a} and \eqref{q4.b}, respectively. Now it remains to reduce the prepotential ${\mathfrak V}_{\a(2s-2)}$ to ${\cN=1}$ AdS superspace, following the same approach as outlined in the type II series. 
The gauge transformation \eqref{3.3} allows us to choose a gauge condition
\bea
{\mathfrak V}_{\a(2s-2)}| =0~. \label{v2}
\eea
The compensator ${\mathfrak V}_{\a(2s-2)}$ is then equivalent to the following real ${\cal N}=1$ superfields, which we define as follows:
\begin{subequations}
\bea
\U_{\b; \, \a(2s-2)} &:=& \ri \bm \nabla^{\underline{2}}_{\b} {\mathfrak V}_{\a(2s-2)}|~, \label{v3} \\
V_{\a(2s-2)} &:=& \frac{\ri}{4}(\bm \nabla^{\underline{2}})^2 {\mathfrak V}_{\a(2s-2)}|~. \label{v4}
\eea
\end{subequations}
The residual gauge freedom, which preserves the gauge condition \eqref{v2} is described by a real unconstrained ${\cal N}=1$ superfield $\eta_{\a(2s-2)}$ defined by
\bea
\x_{\a(2s-2)}| = -\frac{\ri}{2} \eta_{\a(2s-2)}~, \qquad \bar \eta_{\a(2s-2)} = \eta_{\a(2s-2)}~. \label{v5}
\eea
As a result, we may determine how \eqref{v3} and \eqref{v4} vary under $\eta$-transformation
\begin{subequations}
\bea
\d_{\eta} \U_{\b; \, \a(2s-2)} &=& \ri \nabla_{\b} \eta_{\a(2s-2)}~, \label{v6}\\
\d_{\eta} V_{\a(2s-2)} &=&0~.
\eea
\end{subequations}

Next, we analyse the $\l$-gauge transformation and reduce the field strength $\mathbb{V}_{\a(2s-2)}$ to ${\rm AdS}^{3|2}$. 
In the real basis for the covariant derivatives, the real linearity constraint \eqref{3.5} turns into:
\begin{subequations}
\bea
(\bm \nabla^{\underline{2}})^2 \mathbb{V}_{\a(2s-2)} &=& (\bm \nabla^{\underline{1}})^2 \mathbb{V}_{\a(2s-2)}~, \\
\bm \nabla^{\underline{1}\, \b} \bm \nabla^{\underline{2}}_{\b} \mathbb{V}_{\a(2s-2)} &=& 0~.
\eea
\end{subequations}
This tells us that $\mathbb{V}_{\a(2s-2)}$ is equivalent to two real ${\cN=1}$ superfields
\bea
\mathbb{V}_{\a(2s-2)}|~, \qquad \quad \ri \bm \nabla^{\underline{2}}_{\b} \mathbb{V}_{\a(2s-2)}|~.
\eea
The relation between the field strength $\mathbb{V}_{\a(2s-2)}$ and the prepotential ${\mathfrak V}_{\a(2s-2)}$ is given by \eqref{3.4}, which can be expressed as
\bea
\mathbb{V}_{\a(2s-2)} = -\frac{\ri}{2}\Big\{ (\bm \nabla^{\underline{1}})^2 + (\bm \nabla^{\underline{2}})^2\Big\} {\mathfrak V}_{\a(2s-2)}~. 
\label{bbv1}
\eea
We now compute the bar-projection of \eqref{bbv1} in the gauge \eqref{v2} and make use of the definition \eqref{v4} to obtain
\bea
\mathbb{V}_{\a(2s-2)}| = -2 V_{\a(2s-2)}~.
\eea
The bar-projection of $\ri \bm \nabla^{\underline{2}}_{\b} \mathbb{V}_{\a(2s-2)}$ leads to the ${\cal N}=1$ field-strength
\bea
 \O_{\b;\, \a(2s-2)}&:=& \ri \bm \nabla^{\underline{2}}_{\b} \mathbb{V}_{\a(2s-2)}|
 \non \\
 \qquad &=& -\ri \Big( \nabla^{\g} \nabla_{\b} - 4\ri \cS \delta_{\b}\,^{\g}\Big){\U}_{\g; \,\a(2s-2)} ~,
 \label{eqO}
\eea
which is a real superfield, $\O_{\b;\, \a(2s-2)}= \bar \O_{\b;\, \a(2s-2)}$, and is a descendant of the real unconstrained prepotential $\U_{\b; \, \a(2s-2)}$ defined modulo gauge transformation \eqref{v6}. One may check that the field strength is invariant under \eqref{v6} and obeys the condition 
\bea
\nabla^{\b}\O_{\b;\, \a(2s-2)}=0~.
\eea
Let us express the gauge transformation of $\mathbb{V}_{\a(2s-2)}$, eq.~\eqref{3.6} in terms of the real basis for the covariant derivatives. This leads to 
\bea
\d  \mathbb{V}_{\a(2s-2)} &=& -\frac{1}{2s+1} \Big\{ \bm \nabla^{\1 \b} \bm \nabla^{\underline{2}\g}\Big( g_{\b \g \a(2s-2)} - {\bar g}_{\b \g \a(2s-2)}\Big)
\non \\
&+& \bm \nabla^{\b \g}\Big( g_{\b \g \a(2s-2)} + {\bar g}_{\b \g \a(2s-2)}\Big) \Big\}~, 
\eea
One should also express its corollary $\bm \nabla^{\underline{2}}_{\b} \d \mathbb{V}_{\a(2s-2)}$ in the real basis for the covariant derivatives. We determine the gauge transformations law for $V_{\a(2s-2)}$ and $\O_{\b;\, \a(2s-2)}$ to be
\begin{subequations}
\bea
\d V_{\a(2s-2)} &=& \frac{1}{2s} \nabla^{\b} \z_{\b \a(2s-2)}~, \label{N1-Vgauge}\\
\d \O_{\b; \, \a(2s-2)} &=& \frac{1}{2s+1}\big( \nabla^{\g} \nabla_{\b} \nabla^{\d} -4 \ri \cS \nabla^{\d} \d_{\b} \,^{\g} \big) \z_{\d \g \a(2s-2)}~.\label{N1-Ogauge}
\eea
\end{subequations}
From \eqref{N1-Ogauge} we read off the transformation law for the prepotential $\U_{\b;\, \a(2s-2)}$:
\bea
\d \U_{\b; \,\a(2s-2)} &=& \frac{\ri}{2s+1}\Big(\nabla^{\g}\z_{\g \b \a(2s-2)}+ (2s+1) \nabla_{\b} \eta_{\a(2s-2)}\Big)~, \label{N1-Ugauge}
\eea
where we have also taken into account the $\eta$-gauge freedom \eqref{v6}.

Performing ${\cal N}=1$ reduction to the original type III action \eqref{action2}, we arrive at two decoupled ${\cal N}=1$ actions 
\bea
S^{(\rm III)}_{(s+\hf)}[{\mathfrak H}_{\a(2s)} ,{\mathfrak V}_{\a(2s-2)} ] 
= S^{\perp}_{(s+\hf)}[H_{\a(2s+1)} ,{\U}_{\b; \,\a(2s-2)} ] 
+ S^{\parallel}_{(s)}[H_{\a(2s)} ,{V}_{\a(2s-2)} ]~.
\eea
We will present the exact form of these actions in the next subsection. 

%%%%%%%%%%%%%%%%%%%%%%%%%%%%%%%%%%
%%%%%%%%%%%%%%%%%%%%%%%%%%%%%%%%%%%

\subsection{Massless higher-spin ${\cal N}=1$ supermultiplets in AdS${}_3$} \label{ss42}

Upon reduction to ${\cal N}=1$ superspace, the type III theory leads to 
two ${\cal N}=1$ supersymmetric gauge theories. 

%%%%%%%%%%%%%%%%%%%%%%%%%%%%%%%%%

\subsubsection{Longitudinal formulation for massless superspin-$s$ multiplet}

One of the two $\cN=1$ theories provides an off-shell realisation for 
massless superspin-$s$ multiplet described in terms of the real unconstrained superfields
\bea
\cV^{\parallel}_{(s)}= \Big\{H_{\a(2s)}, V_{\a(2s-2)} \Big\} ~,
\label{int}
\eea
which are defined modulo gauge transformations of the form
\begin{subequations} \label{int-gauge}
\bea
\d H_{\a(2s)} &=& \nabla_{(\a_1} \z_{\a_2 \dots \a_{2s})}~, \\
\d V_{\a(2s-2)} &=& \frac{1}{2s} \nabla^{\b} \z_{\b \a(2s-2)}~,
\eea
\end{subequations}
where the gauge parameter $\z_{\a(2s-1)}$ is unconstrained real.
The gauge-invariant action is given by
\bea\label{action-t3}
\lefteqn{S^{\parallel}_{(s)}[H_{\a(2s)} ,V_{\a(2s-2)} ]
= \Big(-\hf \Big)^{s} 
\int 
\rd^{3|2}z
\, E \, \bigg\{
\frac{1}{2} H^{\a(2s)} \big(\ri \de^2  +4 \cS\big)H_{\a(2s)} }
\non \\
&& \qquad 
- \frac{\ri}{2}\de_{\b}H^{\b \a(2s-1)} \de^{\g}H_{\g \a(2s-1)}
-(2s-1) V^{\a(2s-2)} \nabla^{\b \g} H_{\b \g \a(2s-2)}
 \\
&& \qquad +(2s-1)\Big(\hf V^{\a(2s-2)} \big( \ri \nabla^2 +8s\cS\big)V_{\a(2s-2)}
+(s-1) \nabla_{\b}V^{\b \a(2s-3)} \nabla^{\g}V_{\g \a(2s-3)} \Big) 
~.
\non
\eea
Modulo an overall normalisation factor, \eqref{action-t3} coincides with the off-shell ${\cal N}=1$ supersymmetric action for massless superspin-$s$ multiplet in the form given in \cite{KP}. In the flat-superspace limit it reduces to the action derived in 
\cite{KT}.

Although the structure $\nabla_{\b}V^{\b \a(2s-3)} \nabla^{\g}V_{\g \a(2s-3)} $ 
in \eqref{action-t3} is not defined for $s=1$, 
it comes with the factor $(s-1)$  and thus drops out from \eqref{action-t3}
for $s=1$. The resulting gauge-invariant action 
\bea
S^{\parallel}_{(1)}[H_{\a(2)} ,V ]
&=& -\hf 
\int 
\rd^{3|2}z
\, E \, \bigg\{
\frac{1}{2} H^{\a(2)} \big( \ri \de^2 +4\cS\big) H_{\a(2)} 
- \frac{\ri}{2}\de_{\b}H^{\b \a} \de^{\g}H_{\g \a}
\non \\
&& \qquad  
- V \nabla^{\b \g} H_{\b \g}
+\frac{1}{2} V \big( \ri \nabla^2  +8\cS\big) V
\bigg\}
\label{425}
\eea
describes an off-shell massless gravitino multiplet in AdS${}_3$. 
In the flat-superspace limit, it reduces to the gravitino multiplet model described in \cite{Siegel} (see also \cite{KT}).

%%%%%%%%%%%%%%%%%%%%%%%%%%%%%%%%%

\subsubsection{Transverse formulation for massless superspin-$(s+\hf)$ multiplet}

The other theory provides an off-shell formulation for massless  
superspin-$(s+\hf) $ multiplet. It is described by  the unconstrained superfields
\bea
\cV^{\perp}_{(s+\hf )}= \Big\{H_{\a(2s+1)}, \U_{\b; \, \a(2s-2)} \Big\} ~, 
\label{t3new}
\eea
which are defined modulo gauge transformations of the form 
\begin{subequations} \label{t3new-gauge}
\bea
\d H_{\a(2s+1)} &=& \ri \nabla_{(\a_1} \z_{\a_2 \dots \a_{2s+1})}~,\\
\d \U_{\b; \,\a(2s-2)} &=& \frac{\ri}{2s+1}\big( \nabla^{\g} \z_{\g \b \a(2s-2)}+ (2s+1) \nabla_{\b} \eta_{\a(2s-2)}\big)~.
\eea
\end{subequations}

The gauge-invariant  action is 
\begin{subequations}
\label{action-t3-new-complete}
\bea
&&S^{\perp}_{(s+\hf)}[{H}_{(2s+1)} ,\U_{\b; \,\a(2s-2)} ]
= \Big(-\hf \Big)^{s} 
\int \rd^{3|2}z\, E \,\bigg\{-\frac{\ri}{2} H^{\a(2s+1)} {\mathbb{Q}} H_{\a(2s+1)}
% }
\non \\
&& \qquad \qquad -\frac{\ri}{8} \nabla_{\b} H^{\b \a(2s)} \nabla^2 \nabla^{\g}H_{\g \a(2s)}+\frac{\ri}{8}{\nabla}_{\b \g}H^{\b \g \a(2s-1)} {\nabla}^{\rho \d}H_{\rho \d \a(2s-1)}
\non \\
&& \qquad \qquad 
-\frac{\ri}{4}(2s-1) \O^{\b; \,\a(2s-2)} \nabla^{\g \d}H_{\g \d \b \a(2s-2)}
\non\\
&& \qquad \qquad -\frac{\ri}{8}(2s-1)\Big(\O^{\b ;\, \a(2s-2)} \O_{\b ;\, \a(2s-2)}
-2(s-1)\O_{\b;}\,^{\b \a(2s-3)} \O^{\g ;}\,_{\g \a(2s-3)}  \Big) 
\non \\
&& \qquad \qquad 
+ \cS \Big( H^{\a(2s+1)} \big( \nabla^2 - 4\ri \cS\big) H_{\a(2s+1)} 
+ \hf \ \de_{\b}H^{\b \a(2s)} \de^{\g}H_{\g \a(2s)}
\Big)
\non \\
&& \qquad \qquad + \ri s (2s-1) \cS \,\U^{\b ;\, \a(2s-2)} \O_{\b ; \, \a(2s-2)}
\bigg\}~,
\label{action-t3-new}
\eea
where $\O_{\b; \a(2s-2)}$ denotes the real  field strength
\bea
 \O_{\b;\, \a(2s-2)}
  = -\ri \Big( \nabla^{\g} \nabla_{\b} - 4\ri \cS \delta_{\b}\,^{\g}\Big){\U}_{\g; \,\a(2s-2)} ~,
 \qquad \nabla^\b  \O_{\b;\, \a(2s-2)}=0~.
\eea
\end{subequations}
This action defines a new ${\cal N}=1$ supersymmetric higher-spin theory which did not appear in \cite{HKO,KT,KP}.

The structure $\O_{\b;}\,^{\b \a(2s-3)} \O^{\g ;}\,_{\g \a(2s-3)} $
in \eqref{action-t3-new} is not defined for $s=1$. However it comes 
with the factor $(s-1)$ and hence drops out from \eqref{action-t3-new} 
for $s=1$. The resulting gauge-invariant action 
\bea
S^{\perp}_{(\frac 32)}[{H}_{\a(3)} ,\U_{\b} ]
&=& -\hf \int \rd^{3|2}z
\, E \,\bigg\{-\frac{\ri}{2} H^{\a(3)} {\mathbb{Q}} H_{\a(3)} 
  -\frac{\ri}{8} \nabla_{\b} H^{\b \a(2)} \nabla^2 \nabla^{\g}H_{\g \a(2)}
\non \\
&&
 +\frac{\ri}{8}{\nabla}_{\b \g}H^{\b \g \a} {\nabla}^{\rho \d}H_{\rho \d \a}
-\frac{\ri}{4} \O^{\b} \nabla^{\g \d}H_{\g \d \b }
\non \\
&&
+ \cS \Big( H^{\a(3)} \big( \nabla^2 -4\ri \cS\big) H_{\a(3)} 
+ \hf  \de_{\b}H^{\b \a(2)} \de^{\g}H_{\g \a(2)}
\Big)
\non \\
&& 
 -\frac{\ri}{8} \O^{\b  } \O_{\b }
+ \ri  \cS \,\U^{\b } \O_{\b }\bigg\}
\eea
provides an off-shell formulation for a linearised supergravity 
multiplet in AdS${}_3$. In the flat-superspace limit, it reduces to the linearised
supergravity model proposed in \cite{KT}.

%%%%%%%%%%%%%%%%%%%%%%%%%%%%%%%%%%%
%%%%%%%%%%%%%%%%%%%%%%%%%%%%%%%%%%%%%

\section{Analysis of the results} \label{section5}

Let $s>0$ be a positive integer. For each superspin value, 
integer $(s)$ or half-integer $(s+\hf)$, we have constructed two off-shell 
formulations which have been called longitudinal and transverse. 
Now we have to explain this terminology.

Consider a field theory in AdS${}^{3|2}$ that is described in terms of a real tensor 
superfield $V_{\a(n)}$.  We assume the action to have the form
\bea
S^{\parallel} [ V_{\a(n)} ]= \int \rd^{3|2}z \, E\, \cL \big(\ri^{n+1}  \nabla_\b V_{\a(n)} \big)~.
\label{long5.1}
\eea 
It is natural to call $\nabla_\b V_{\a(n)} $ a longitudinal superfield, 
by analogy with a longitudinal vector field.
This theory possesses a dual formulation that is obtained by introducing 
a first-order action 
\bea
S_{\text{first-order}} = \int \rd^{3|2}z \, E\, \Big\{ \cL \big( \S_{\b; \, \a(n)} \big)
+ \ri^{n+1} \cW^{\b; \, \a(n) }  \S_{\b; \,\a(n)} \Big\}~,
\label{first-order}
\eea
where $\S_{\b;\a(n)} $ is unconstrained and the Lagrange multiplier is 
\bea
 \cW_{\b; \,\a(n)} =
  \ri^{n+1} \Big( \nabla^{\g} \nabla_{\b} - 4\ri \cS \delta^{\g}_{\,\b}\Big){\Psi}_{\g; \,\a(n)} ~,
  \qquad \nabla^\b \cW_{\b; \,\a(n)} =0~,
  \label{W5.3}
 \eea
for some unconstrained prepotential ${\Psi}_{\g; \,\a(n)} $. 
Varying \eqref{first-order} with respect to ${\Psi}_{\g; \,\a(n)} $ gives
\bea
\nabla^\b \nabla_\g \S_{\b; \,\a(n)} - 4\ri \cS \S_{\g; \,\a(n)} =0 
\quad \implies \quad \S_{\b; \,\a(n)} = \ri^{n+1} \nabla_\b V_{\a(n)}~,
\eea
and then $S_{\text{first-order}} $ reduces to the original action \eqref{long5.1}.
On the other hand, we may start from $S_{\text{first-order}}$ 
and integrate $\S_{\b;\a(n)} $ out.  This will lead to a dual action of the form 
\bea
S^{\perp} [ {\Psi}_{\g; \,\a(n)} ]= \int \rd^{3|2}z \, E\,  
\cL_{\rm dual} \big(  \cW_{\b; \, \a(n) } \big)~.
\label{tran5.5}
\eea
This is a gauge theory since the action is invariant under 
gauge transformations
\bea
\d {\Psi}_{\g; \,\a(n)} = \ri^{n+1} \nabla_\g \eta_{\a(n)} ~.
\eea
The gauge-invariant field strength $\cW_{\b; \, \a(n) } $ can be called a transverse superfield, due to the constraint \eqref{W5.3} it obeys. 

It is natural to call the dual formulations \eqref{long5.1} and \eqref{tran5.5} 
as longitudinal and transverse, respectively.

Now,  let us consider the transverse and longitudinal  formulations 
for the massless superspin-$s$ models, which 
are given by eqs. \eqref{action-t2-new} and \eqref{action-t3}, respectively. 
These actions depend parametrically on $\cS$, the curvature of AdS superspace.
We denote by $S^{\perp}_{(s)}[H_{\a(2s)} ,{\Psi}_{\b; \,\a(2s-2)}]_{\rm FS}$
and $S^{\parallel}_{(s)}[H_{\a(2s)} ,V_{\a(2s-2)} ]_{\rm FS}$
these actions in the limit $\cS=0$, which corresponds to a flat superspace.
The dynamical systems  $S^{\perp}_{(s)}[H_{\a(2s)} ,{\Psi}_{\b; \,\a(2s-2)}]_{\rm FS}$
and $S^{\parallel}_{(s)}[H_{\a(2s)} ,V_{\a(2s-2)} ]_{\rm FS}$ prove to 
be related to each other by  the Legendre transformation described above.
Thus $S^{\perp}_{(s)}[H_{\a(2s)} ,{\Psi}_{\b; \,\a(2s-2)}]_{\rm FS}$
and $S^{\parallel}_{(s)}[H_{\a(2s)} ,V_{\a(2s-2)} ]_{\rm FS}$ are dual formulations of the same theory. This duality does not survive if $\cS$ is non-vanishing.

The same feature characterises the longitudinal and transverse  formulations 
for the massless superspin-$(s+\hf)$ multiplet, which 
are described by the actions \eqref{action-t2-half} and \eqref{action-t3-new-complete}, respectively. The flat-superspace counterparts of these higher-spin models, which we denote by
$S^{\parallel}_{(s+\hf)}[H_{\a(2s+1)} ,L_{\a(2s-2)} ]_{\rm FS}$ and 
$S^{\perp}_{(s+\hf)}[{H}_{\a(2s+1)} ,\U_{\b; \,\a(2s-2)} ]_{\rm FS}$,
are dual to each other. However, this duality does not survive 
if we turn on a non-vanishing AdS curvature.

The above discussion can be illustrated by considering 
the model for linearised gravity  in AdS${}_3$. It is described by the action 
\begin{align}
S_{\rm gravity}=&\frac 18
\int\text{d}^3x\, e \,
\bigg\{{\mathfrak h}^{\a(4)}\Box {\mathfrak h}_{\a(4)}
- \nabla_{\b(2)}{\mathfrak h}^{\b(2)\a(2)}\nabla^{\g(2)}
{\mathfrak h}_{\a(2)\g(2)}\notag\\
&+\frac{1}{2}\nabla^{\a(2)} {\mathfrak y} \nabla^{\b(2)}
{\mathfrak h}_{\a(2)\b(2)}
- \frac 14 \nabla^{\a(2)} {\mathfrak y} \nabla_{\a(2)} {\mathfrak y}
+8\mathcal{S}^2{\mathfrak h}^{\a(4)}{\mathfrak h}_{\a(4)}
+6\mathcal{S}^2{\mathfrak y}^2
\bigg\}~,
\end{align}
which is invariant under gauge transformations 
\bea
\d_{\x} {\mathfrak h}_{\a(4)}=\nabla_{(\a_1\a_2}\zeta_{\a_3\a_4)}~,\qquad
\d_{\x} {\mathfrak y}=\frac{2}{3}\nabla^{\a(2)}\z_{\a(2)}~.
\eea
In the flat-space limit, $\cS=0$, the model possesses a dual formulation in which the scalar compensator $\mathfrak y$ is replaced with a gauge one-form.\footnote{There is another dual realisation in which ${\mathfrak h}_{\a(4)}$ turns into 
a gauge one-form ${\mathfrak h}_{b;\a(4)}$ with an additional gauge freedom.}
However, such a duality transformation cannot be lifted to AdS${}_3$.

%%%%%%%%%%%%%%%%%%%%%%%%%%%%%%%%
%%%%%%%%%%%%%%%%%%%%%%%%%%%%%%%%

\section{Non-conformal higher spin supercurrents} \label{section6}

In the previous sections, we have shown that there exist
two different off-shell formulations for the massless higher-spin ${\cal N}=1$ supermultiplets. Massless half-integer superspin theory can be realised in terms of the dynamical variables \eqref{hf} and \eqref{t3new}, while the models \eqref{t2new} and \eqref{int} define massless multiplet of integer superspin $s$, with $s >1$. These models lead to different ${\cal N}=1$ higher-spin supercurrent multiplets. Our aim in this section is to describe the general structure of ${\cal N}=1$ supercurrent multiplets in AdS. 

%%%%%%%%%%%%%%%%%%%%%%%%%%%%%%%%%%
%%%%%%%%%%%%%%%%%%%%%%%%%%%%%%%%%

\subsection{${\cal N}=1$ supercurrents: Half-integer superspin case}

Our half-integer supermultiplet in the longitudinal formulation \eqref{hf} can be coupled to external sources
\bea
S^{(s+\hf)}_{\rm source}= \int \rd^{3|2}z \, E\, \Big\{ 
{\ri} {H}^{ \a (2s+1)} J_{ \a (2s+1)}
+ 4{L}^{ \a (2s-2)} {S}_{ \a (2s-2)}
 \Big\}~.
\eea
The condition that the above action is invariant under the gauge transformations \eqref{hf-gauge} gives the conservation equation
\bea
\de^{\b}J_{\b \a(2s)} = -\frac{2s}{(2s+1)} \de_{(\a_1 \a_2} S_{\a_3 \cdots \a_{2s})}~.
\label{ce-hf11}
\eea

For the transverse theory \eqref{t3new} described by the prepotentials $\{ H_{\a(2s+1)},\U_{\b;\, \a(2s-2)}\}$, we construct an action functional of the form
\bea
S^{(s+\hf)}_{\rm source}= \int \rd^{3|2} z
\, E\, \Big\{ 
{\ri} {H}^{ \a (2s+1)} J_{ \a (2s+1)}
+ 2 \ri s\, {\U}^{\b; \, \a (2s-2)} {U}_{ \b; \, \a (2s-2)}
 \Big\}~.
\eea
Requiring that the action is invariant under the gauge transformations \eqref{t3new-gauge} leads to
\bea \label{ce-hf12}
\de^{\b}J_{\b \a(2s)} = \frac{2s}{2s+1} \de_{(\a_1} U_{\a_2 \cdots \a_{2s})}~, \quad 
\de^{\b} U_{\b; \, \a(2s-2)} = 0~. 
\eea
From the above consideration, it follows that the most general conservation equation in the half-integer superspin case takes the form
\bsubeq \label{ce-hf13}
\bea
\de^{\b}J_{\b \a(2s)} &=& \frac{2s}{2s+1} \bigg(\de_{(\a_1} U_{\a_2 \cdots \a_{2s})} 
- \de_{(\a_1 \a_2} S_{\a_3 \cdots \a_{2s})} \bigg)~, \label{ce-hf13a} \\
\de^{\b} U_{\b; \, \a(2s-2)} &=& 0~.\label{ce-hf13b}
\eea
\esubeq

%%%%%%%%%%%%%%%%%%%%%%%%%%%%%%%
%%%%%%%%%%%%%%%%%%%%%%%%%%%%%%%

\subsection{${\cal N}=1$ supercurrents: Integer superspin case}
In complete analogy with the half-integer superspin case, we couple the prepotentials \eqref{int} in terms of which the integer superspin-$s$ is described, to external sources 
\bea
S^{(s)}_{\rm source}= \int \rd^{3|2}z \, E\, \Big\{ 
 {H}^{ \a (2s)} J_{ \a (2s)}
+ 2s \,{V}^{ \a (2s-2)} {R}_{ \a (2s-2)}
 \Big\}~.
\eea
For such an action to be invariant under the gauge freedom \eqref{int-gauge}, the sources must be conserved 
\bea
\de^{\b}J_{\b \a(2s-1)} = \de_{(\a_1} R_{\a_2 \cdots \a_{2s-1})}~.
\label{ce-int11}
\eea

Next, we turn to the transverse formulation \eqref{t2new} characterised by the prepotentials $\{ H_{\a(2s+1)},\J_{\b;\, \a(2s-2)}\}$ and construct an action functional
\bea
S^{(s)}_{\rm source}= \int \rd^{3|2}z
\, E\, \Big\{ 
{H}^{ \a (2s)} J_{ \a (2s)}
+ \ri {\J}^{\b; \, \a (2s-2)} {T}_{ \b; \, \a (2s-2)}
 \Big\}~.
\eea
Demanding that the action be invariant under the gauge transformations \eqref{t2new-gauge}, we derive the following conditions
\bea
\de^{\b}J_{\b \a(2s-1)} = \ri\, T_{\a(2s-1)}~, \qquad \de^{\b} T_{\b; \, \a(2s-2)} = 0~. \label{ce-int12}
\eea
From the above consideration, the most general conservation equation for the multiplet of currents in the integer superspin case is given by
\bsubeq \label{ce-int13}
\bea
\de^{\b}J_{\b \a(2s-1)} &=&  \de_{(\a_1} R_{\a_2 \cdots \a_{2s-1})}+ 
\ri T_{\a(2s-1)}~,\\
\de^{\b} T_{\b; \, \a(2s-2)} &=& 0~. 
\eea
\esubeq

%%%%%%%%%%%%%%%%%%%%%%%%%%%%%%%%%%
%%%%%%%%%%%%%%%%%%%%%%%%%%%%%%%%%%

\subsection{From ${\cal N}=2$ supercurrents to $\cN=1$ supercurrents}

In our recent paper \cite{HK18}, we constructed the general conservation equation for the ${\cal N}=2$ higher-spin supercurrent multiplets in (2,0) AdS superspace, which takes the form
\bea
\label{ce1}
\bar \cD^{\b} \mathbb{J}_{\b \a(2s-1)} = \bar \cD_{(\a_1} \big( \mathbb{Y}_{\a_2 \dots \a_{2s-1})} + \ri \mathbb{Z}_{\a_2 \dots \a_{2s-1})} \big)~.
\eea
Here $\mathbb{J}_{\a(2s)}$ denotes the higher-spin supercurrent, while the trace supermultiplets $\mathbb{Y}_{\a(2s-2)} $ and $\mathbb{Z}_{\a(2s-2)} $
are covariantly linear. The explicit form of this multiplet of currents was presented by considering simple ${\cal N}=2$ supersymmetric models for a chiral scalar superfield. 
Unlike in 4D ${\cal N}=1$ supergravity where every supersymmetric matter theory can be coupled to only one of the off-shell supergravity formulations (either old-minimal or new-minimal), here in the (2,0) AdS case our trace multiplets require both type II and type III compensators to couple to. 

The general conservation equation \eqref{ce1} naturally gives rise to the ${\cal N}=1$ higher-spin supercurrent multiplets discussed in the previous subsection. One may show that in the real basis, \eqref{ce1} turns into:
\bsubeq \label{ce2}
\bea
{\bm \de^{\1 \b}} \mathbb{J}_{\b \a(2s-1)} &=& \bm \de^{\1}_{(\a_1} \mathbb{Y}_{\a_2 \cdots \a_{2s-1})} - \bm \de^{\2}_{(\a_1} \mathbb{Z}_{\a_2 \cdots \a_{2s-1})}~, \\
\bm \de^{\2 \b} \mathbb{J}_{\b \a(2s-1)} &=& \bm \de^{\1}_{(\a_1} \mathbb{Z}_{\a_2 \cdots \a_{2s-1})} + \bm \de^{\2}_{(\a_1} \mathbb{Y}_{\a_2 \cdots \a_{2s-1})}~,
\eea
\esubeq
The real linearity constraints on the trace supermultiplets are equivalent to
\begin{subequations} \label{ce3}
\bea
(\bm \nabla^{\underline{2}})^2 \mathbb{Y}_{\a(2s-2)} &=& (\bm \nabla^{\underline{1}})^2 \mathbb{Y}_{\a(2s-2)}~, \qquad 
\bm \nabla^{\underline{1}\, \b} \bm \nabla^{\underline{2}}_{\b} \mathbb{Y}_{\a(2s-2)} = 0~, \\
(\bm \nabla^{\underline{2}})^2 \mathbb{Z}_{\a(2s-2)} &=& (\bm \nabla^{\underline{1}})^2 \mathbb{Z}_{\a(2s-2)}~, \qquad 
\bm \nabla^{\underline{1}\, \b} \bm \nabla^{\underline{2}}_{\b} \mathbb{Z}_{\a(2s-2)} = 0~.
\eea
\end{subequations}

It follows from \eqref{ce2} and \eqref{ce3} that $\mathbb{J}_{\a(2s)}$ contains two independent real ${\cal N}=1$ supermultiplets:
\bsubeq \label{J}
\bea
J_{\a(2s)} &:=& \mathbb{J}_{\a(2s)}|~, \\
J_{\a_(2s+1)} &:=& \ri \bm \de^{\2}_{(\a_1} \mathbb{J}_{\a_2 \cdots \a_{2s+1})}|~,
\eea
\esubeq
while the independent real ${\cal N}=1$ components of $\mathbb{Y}_{\a(2s-2)}$ and $\mathbb{Z}_{\a(2s-2)}$ are defined by
\bsubeq \label{trace}
\bea
R_{\a(2s-2)} &:=& \mathbb{Y}_{\a(2s-2)}|~, \qquad U_{\b;\, \a(2s-2)}:= \ri \bm \de^{\2}_{\b} \mathbb{Y}_{\a(2s-2)}|~, \\
S_{\a(2s-2)} &:=& \mathbb{Z}_{\a(2s-2)}|~, \qquad T_{\b;\, \a(2s-2)}:= \ri \bm \de^{\2}_{\b} \mathbb{Z}_{\a(2s-2)}|~.
\eea
\esubeq
Making use of \eqref{ce3}, one may readily show that 
\bsubeq
\bea
\de^{\b} U_{\b; \, \a(2s-2)} = 0~, \label{ce-hf1} \\
\de^{\b} T_{\b; \, \a(2s-2)} = 0~. \label{ce-int1} 
\eea
\esubeq
On the other hand, eq.~\eqref{ce2} implies that the ${\cal N}=1$ superfields obey the following conditions
\bsubeq
\bea
&&\de^{\b}J_{\b \a(2s)} = \frac{2s}{2s+1} \Big(\de_{(\a_1} U_{\a_2 \cdots \a_{2s})} 
- \de_{(\a_1 \a_2} S_{\a_3 \cdots \a_{2s})} \Big)~, \label{ce-hf2}\\
&&\de^{\b}J_{\b \a(2s-1)} =  \de_{(\a_1} R_{\a_2 \cdots \a_{2s-1})}+ 
\ri T_{\a(2s-1)}~.  \label{ce-int2}
\eea
\esubeq
Indeed, the right-hand side of eq.~\eqref{ce-hf2} coincides with \eqref{ce-hf13a}. Therefore, eqs.~\eqref{ce-hf1} and \eqref{ce-hf2} define the ${\cal N}=1$ higher-spin current multiplets associated with the massless half-integer superspin formulations \eqref{hf} and \eqref{t3new}. In a similar way, it can be observed that eqs.~\eqref{ce-int1} and \eqref{ce-int2} correspond to the ${\cal N}=1$ higher-spin supercurrents for the two integer superspin models  \eqref{t2new} and \eqref{int}.

%%%%%%%%%%%%%%%%%%%%%%%%%%%%%%%%%%%
%%%%%%%%%%%%%%%%%%%%%%%%%%%%%%%%%%%

\section{Examples of ${\cal N}=1$ higher-spin supercurrents} \label{section7}
In this section we give an explicit realisation of the ${\cal N}=1$ multiplet of higher-spin supercurrent introduced earlier.

Consider a massless chiral scalar multiplet in (2,0) AdS superspace with action \cite{HK18}
\bea 
S = \int \rd^3x \,\rd^2 \q  \,\rd^2 \bar \q
\, {\bm E} \,\bar \F \F ~, \qquad \bar \cD_{\a} \F =0~.
\label{ch-action}
\eea
The chiral superfield is charged under  the $R$-symmetry group $\rm U(1)_R$,
\bea
J \F = -r \F~.
\eea
This action is a special case of the supersymmetric nonlinear sigma model studied in subsection \eqref{ss2.3} with a vanishing superpotential, $W(\F)=0$~. 
Making use of \eqref{N1-sigma}, the reduction of the action 
\eqref{ch-action} to $\cN=1$ AdS superspace is
\bea
S &=& \int \rd^{3|2}z\,E \,
\Big\{-\ri \de^{\a} \bar \vf \de_{\a}  \vf + 4 r \cS\, \bar \vf \vf \Big\}~, 
\eea
where we have denoted $\vf:= \F|$~. This action is manifestly ${\cal N}=1$ supersymmetric. It also possesses hidden second supersymmetry and $\rm U(1)_R$ invariance. These transformations are
\bea
\d_{\e} \vf = \ri \e^{\a} \de_{\a}\vf -{\ri}\e \, r \vf ~, \qquad
\d_{\e} \bar \vf = \ri \e^{\a} \de_{\a} \bar \vf 
+{\ri}\e \,r {\bar \vf}~,
\eea
where $\e^\a$ is given in terms or $\e$ according to \eqref{u1-c-a},
and the real parameter $\e$ is constrained by \eqref{u1-c-b}.
It can be shown that on the mass shell it holds that
\bea 
(\ri \de^2 +4 r \cS)\vf = 0~, \qquad (\ri \de^2 +4 r \cS)\bar \vf = 0~.
\label{eom}
\eea

It was shown in \cite{HK18} that by using the massless equations of motion, $\cD^2 \F=0$, the ${\cal N}=2$ higher-spin supercurrent multiplet associated with the theory \eqref{ch-action} is described by the conservation equation
\bsubeq 
\bea
\cD_{(-1)} \mathbb{J}_{(2s)} &=&  {\cD}_{(1)} {\mathbb T}_{(2s-2)}~.
\eea
Here the real supercurrent $ \mathbb{J}_{(2s)} = \bar {\mathbb{J}}_{(2s)}$ is given by
\bea
\mathbb{J}_{(2s)} &=& \sum_{k=0}^s (-1)^k
\left\{ \hf \binom{2s}{2k+1} 
{\cD}^k_{(2)} \bar \cD_{(1)} \bar \F \,\,
{\cD}^{s-k-1}_{(2)} \cD_{(1)} \F  
+ \binom{2s}{2k} 
{\cD}^k_{(2)} \bar \F \,\, {\cD}^{s-k}_{(2)} \F \right\}~,~~~
\eea
while the trace multiplet $\mathbb{T}_{(2s-2)}$ has the form
\bea
{\mathbb T}_{(2s-2)}&=& 2\ri {\cS}(1-2r)(2s+1)(s+1) \sum_{k=0}^{s-1}\frac{1}{2s-2k+1} (-1)^{k} \binom{2s}{2k+1}
\non \\ 
&& 
\times {\cD}^k_{(2)} \bar \F \,\,{\cD}^{s-k-1}_{(2)} \F ~.
\label{7.6a}
\eea
One may check that  ${\mathbb T}_{(2s-2)}$ is covariantly linear,
\bea
\bar \cD^2 \mathbb{T}_{(2s-2)} =0~, \qquad \cD^2 \mathbb{T}_{(2s-2)} =0~.
\eea
\esubeq
As is seen from \eqref{7.6a}, ${\mathbb T}_{(2s-2)}$ vanishes for  $r =1/2$, 
in which case $\F$ is an $\cN=2$  superconformal multiplet.

The complex trace multiplet $\mathbb{T}_{(2s-2)}$  
may be split into its real and imaginary parts:
\bsubeq
\bea
\mathbb{T}_{(2s-2)} = \mathbb{Y}_{(2s-2)}- \ri \mathbb{Z}_{(2s-2)}~,
\eea
with 
\bea
\mathbb{Y}_{(2s-2)} &=& 2\ri {\cS}(1-2r)(2s+1)(s+1) \sum_{k=0}^{s-1}\frac{2k-s+1}{(2k+3)(2s-2k+1)} 
\non \\ 
&& 
\times (-1)^{k} \binom{2s}{2k+1} {\cD}^k_{(2)} \bar \F \,\,{\cD}^{s-k-1}_{(2)} \F ~,\\
\mathbb{Z}_{(2s-2)} &=& -2 {\cS}(1-2r)(2s+1)(s+1)(s+2) \sum_{k=0}^{s-1}\frac{1}{(2k+3)(2s-2k+1)}
\non \\ 
&& 
\times (-1)^{k} \binom{2s}{2k+1} {\cD}^k_{(2)} \bar \F \,\,{\cD}^{s-k-1}_{(2)} \F ~.
\eea
\esubeq

Note that we make use of a condensed notation \cite{HK18} throughout this section. In this notation, we introduce
auxiliary real variables $\z^\a \in {\mathbb R}^2$, such that any tensor superfield $U_{\a(m)}$ can be associated with the following  field
\bea
U_{(m)} (\z):= \z^{\a_1} \dots \z^{\a_m} U_{\a_1 \dots \a_m}~,
\eea
which is a homogeneous polynomial of degree $m$ in $\z^\a$.
Let us introduce operators which increase the degree 
of homogeneity in $\z^\a$: 
\bea
{\cD}_{(1)} := \z^\a \cD_\a~, \quad 
{\bar \cD}_{(1)} :=  \z^\a \bar \cD_\a~,\quad 
{\cD}_{(2)} := \ri \z^\a \z^\b \cD_{\a\b}~.
\eea
In addition, we also have two operators that decrease the degree 
of homogeneity in $\z^\a$
\bea
\cD_{(-1)} := \cD^\a \frac{\pa}{\pa \z^\a}~, \qquad 
\bar \cD_{(-1)} := \bar \cD^\a \frac{\pa}{\pa  \z^\a}~.
\eea
The operators associated with the real spinor covariant derivatives may be defined in a similar way:
\bea
{\bm \de}^{\1}_{(1)} &:=& \z^\a \bm \de^{\1}_\a , \quad 
{\bm \de}^{\2}_{(1)} :=  \z^\a \bm \de^{\2}_\a~,\quad 
{\bm \de}_{(2)} := \ri \z^\a \z^\b \bm \de_{\a\b}~, \\
{\bm \de}^{\1}_{(-1)} &:=& \bm \de^{\1 \a} \frac{\pa}{\pa \z^\a}~.
\eea

In accordance with \eqref{J}, the supercurrent $\mathbb{J}_{(2s)}$ reduces to two different multiplets upon projection to ${\cal N}=1$ superspace:
\bsubeq 
\bea
J_{(2s)} &:=& \mathbb{J}_{(2s)}\big| \non\\
&=& \sum_{k=0}^s (-1)^{k+1}
\bigg\{ \binom{2s}{2k+1} 
{\de}^k_{(2)} \de_{(1)} \bar \vf \,\,
{\de}^{s-k-1}_{(2)} \de_{(1)} \vf  \non\\
&&- \binom{2s}{2k} 
{\de}^k_{(2)} \bar \vf \,\, {\de}^{s-k}_{(2)} \vf \bigg\}~, \\
J_{(2s+1)} &:=& \ri \bm \de^{\2}_{(1)}\mathbb{J}_{(2s)} \big| =
-\frac{1}{\sqrt{2}}\big(\cD_{(1)} + \bar \cD_{(1)} \big) \mathbb{J}_{(2s)}\big|~,\non\\
&=& (2s+1)\sum_{k=0}^s
\frac{1}{2s-2k+1} (-1)^{k+1} \binom{2s}{2k} \bigg\{
{\de}^k_{(2)} \bar \vf \,\, {\de}^{s-k}_{(2)} \de_{(1)} \vf \non\\
&&+ (-1)^{s-1} {\de}^k_{(2)}\vf \,\,
{\de}^{s-k}_{(2)} \de_{(1)} \bar \vf \bigg\}~,
\eea
\esubeq
of which the former corresponds to the integer superspin current and the latter half-integer superspin current. 

In the case of half-integer superspin, the conservation equation \eqref{ce-hf13} is satisfied provided we impose \eqref{eom}:
\bsubeq
\bea
\de_{(-1)} J_{(2s+1)} = \frac{2s}{2s+1} \bigg( \de_{(1)} U_{(2s-1)}+ \ri \de_{(2)} S_{(2s-2)}\bigg)~,\quad
\de^{\b} U_{\b; \,(2s-2)}= 0~,
\eea
with 
\bea
S_{(2s-2)} &:=& \mathbb{Z}_{(2s-2)} \big| \non\\
&=& -2 {\cS}(1-2r)(2s+1)(s+1)(s+2) \sum_{k=0}^{s-1}\frac{1}{(2k+3)(2s-2k+1)}
\non \\ 
&& 
\times (-1)^{k} \binom{2s}{2k+1} {\de}^k_{(2)} \bar \vf \,\,{\de}^{s-k-1}_{(2)} \vf ~,\\
U_{\b; \,(2s-2)} &:=& -\frac{1}{\sqrt{2}}\big(\cD_{\b} + \bar \cD_{\b} \big) \mathbb{Y}_{(2s-2)}\big|~,\non\\
&=& -2\ri {\cS}(1-2r)(2s+1)(s+1) \sum_{k=0}^{s-1}\frac{2k-s+1}{(2k+3)(2s-2k+1)} 
(-1)^{k} \binom{2s}{2k+1} \non\\
&& \times \bigg\{{\de}^k_{(2)} \bar \vf \,\,{\de}^{s-k-1}_{(2)} \de_{\b} \vf + (-1)^{s+1} {\de}^k_{(2)} \vf \,\,{\de}^{s-k-1}_{(2)} \de_{\b} \bar \vf \non\\
&&\qquad + 2\ri \cS (s-k-1) \z_{\b} \bigg({\de}^k_{(2)} \bar \vf \,\,{\de}^{s-k-2}_{(2)} \de_{(1)} \vf \non\\
&&\qquad+ (-1)^{s+1} {\de}^k_{(2)} \vf \,\,{\de}^{s-k-2}_{(2)} \de_{(1)} \bar \vf \bigg) \bigg\}~.
\eea
\esubeq

It may also be verified that the ${\cal N}=1$ supercurrent multiplet for integer superspin obeys the conditions \eqref{ce-int13} on-shell:
\bsubeq
\bea
\de_{(-1)} J_{(2s)} = \de_{(1)} R_{(2s-2)}+ \ri T_{(2s-1)}~, \quad \de^{\b}T_{\b; \,(2s-2)}=0~.
\eea
with
\bea
R_{(2s-2)} &:=& \mathbb{Y}_{(2s-2)} \big| \non\\
&&= 2\ri {\cS}(1-2r)(2s+1)(s+1)\sum_{k=0}^{s-1}\frac{2k-s+1}{(2k+3)(2s-2k+1)} 
\non \\ 
&& 
\times (-1)^{k} \binom{2s}{2k+1} {\de}^k_{(2)} \bar \vf \,\,{\de}^{s-k-1}_{(2)} \vf ~,\\
T_{\b; \,(2s-2)} &:=& -\frac{1}{\sqrt{2}}\big(\cD_{\b} + \bar \cD_{\b} \big) \mathbb{Y}_{(2s-2)}\big|~,\non\\
&=& 2 {\cS}(1-2r)(2s+1)(s+1)(s+2) \sum_{k=0}^{s-1}\frac{1}{(2k+3)(2s-2k+1)} 
(-1)^{k} \binom{2s}{2k+1} \non\\
&& \times \bigg\{{\de}^k_{(2)} \bar \vf \,\,{\de}^{s-k-1}_{(2)} \de_{\b} \vf + (-1)^{s} {\de}^k_{(2)} \vf \,\,{\de}^{s-k-1}_{(2)} \de_{\b} \bar \vf \non\\
&&\qquad + 2\ri \cS (s-k-1) \z_{\b} \bigg({\de}^k_{(2)} \bar \vf \,\,{\de}^{s-k-2}_{(2)} \de_{(1)} \vf \non\\
&&\qquad+ (-1)^{s} {\de}^k_{(2)} \vf \,\,{\de}^{s-k-2}_{(2)} \de_{(1)} \bar \vf \bigg) \bigg\}~.
\eea
\esubeq

The above technique can also be used to construct ${\cal N}=1$ higher-spin supercurrents for the Abelian vector multiplets model described by the action \eqref{2.46}. We will not elaborate on a construction in the present work. 

In four dimensions, various aspects of the 
higher-spin  supercurrent multiplets were studied in 
\cite{KMT,BGK1,HK1,HK2,KKvU} in the $\cN=1$ super-Poincar\'e case
and in \cite{BHK} for $\cN=1$ AdS supersymmetry. In particular, 
the general non-conformal higher-spin supercurrent multiplets
for $\cN=1$ supersymmetric field theories in Minkowski space were proposed
in \cite{HK1,HK2}, and their AdS counterparts were formulated in \cite{BHK}.
Explicit realisations of the higher-spin supercurrents were 
derived in \cite{BHK} for various $\cN=1$ supersymmetric theories  in AdS${}_4$, 
including a model of $N$ massive chiral scalar
superfields with an arbitrary mass matrix.

%%%%%%%%%%%%%%%%%%%%%%%%%%%%
%%%%%%%%%%%%%%%%%%%%%%%%%%%%%%%

\section{Applications and open problems}

Let us briefly summarise the main results obtained in this paper. In section \ref{section2} we developed a formalism to reduce every field theory 
with (2,0) AdS supersymmetry to ${\cal N}=1$ AdS superspace. 
In sections \ref{section3} and \ref{section4} we applied this reduction procedure 
to the off-shell massless higher-spin supermultiplets in AdS$_{(3|2,0)}$ constructed in \cite{HK18}. 
For each superspin value, integer ($s$) or half-integer $(s+\hf)$, the reduction produced two off-shell gauge formulations, longitudinal and transverse, for massless ${\cal N}=1$ supermultiplets in AdS$_{3}$. The transverse higher-spin formulations
for massless ${\cal N}=1$ supermultiplets in AdS$_{3}$ are  new gauge theories.
 In section \ref{section5}, we proved that for each superspin value the longitudinal and transverse theories  are dually equivalent only in the flat superspace limit. 
In section \ref{section6}  we formulated, for the first time, the non-conformal higher-spin supercurrent  in ${\cal N}=1$ AdS superspace. 
In \ref{section7} we provided the explicit examples of these supercurrents
for models of a chiral scalar superfield. 

There are several interesting applications of the results obtained in this paper. 
In particular, the massless higher-spin $\cN=1$ supermultiplets in AdS${}_3$, 
which were derived in sections \ref{section3} and \ref{section4}, can be used 
to construct off-shell topologically massive supermultiplets in AdS${}_3$
by extending the approaches advocated in \cite{KO,KT,KP}. Such a massive 
supermultiplet is described by  a gauge-invariant action being the sum of 
 massless and  superconformal higher-spin actions, 
following the philosophy of topologically massive theories
 \cite{Siegel,Schonfeld,DJT1,DJT2}. 

Given a positive integer $n$, the conformal superspin-$\frac{n}{2}$ 
action \cite{K16,KT,KP} is 
\bea
{S}_{\rm{SCS}}^{(n/2)} [ H_{\a(n)}] 
= - \frac{\ri^n}{2^{\left \lfloor{n/2}\right \rfloor +1}}
   \int \rd^{3|2}z \, E\,
 {H}^{\a(n)} 
{W}_{\a(n) }( {H}) ~, 
\label{8.1}
\eea
where $W_{\a (n)}( H)$ denotes the higher-spin super-Cotton tensor. 
The latter is a unique  descendant of $H_{ \a(n) }$ with the properties
\begin{subequations}
\bea
W_{\a(n)} \big(\d_\z H\big)  &=&0~,\\
\nabla^\b W_{\b \a (n-1)} &=&0~,
\label{8.2b}
\eea
\end{subequations}
where $\d_\z H_{\a(n)} $ is the gauge transformation \eqref{3.299}.
These properties imply the gauge invariance of \eqref{8.1}.
In a flat superspace, ${W}_{\a(n) }$ has the form
\cite{K16}
\bea
\cS =0 \quad \implies \quad 
W_{\a_1 \dots \a_n} = 
\Big( -\frac{\ri}{2}\Big)^n
\nabla^{\b_1} \nabla_{\a_1} \dots \nabla^{\b_n} \nabla_{\a_n} H_{\b_1 \dots \b_n}
~.
\eea
The construction of $W_{\a(n)}$ in arbitrary conformally flat backgrounds 
is described in \cite{KP2}.

Given a positive integer $s$, 
there are two off-shell gauge-invariant formulations for a topologically 
massive superspin-$s$ multiplet in AdS${}_3$. The corresponding actions are:
\begin{subequations} \label{8.4}
\bea
S^{\parallel}_{(s)}[H_{\a(2s)} ,V_{\a(2s-2)} |\m]
&=& {S}_{\rm{SCS}}^{(s)} [ H_{\a(2s)}] 
+\m^{2s-1}S^{\parallel}_{(s)}[H_{\a(2s)} ,V_{\a(2s-2)} ]~, \label{8.4a} \\
S^{\perp}_{(s)}[H_{\a(2s)} ,{\Psi}_{\b; \,\a(2s-2)} |\m]
&=& {S}_{\rm{SCS}}^{(s)} [ H_{\a(2s)}] 
+\m^{2s-1}S^{\perp}_{(s)}[H_{\a(2s)} ,{\Psi}_{\b; \,\a(2s-2)}]~. \label{8.4b}
\eea
\end{subequations}
The dynamical system  \eqref{8.4a} was introduced in \cite{KP}, while its flat-superspace 
counterpart appeared earlier in \cite{KT}. The other theory, 
eq.  \eqref{8.4b}, is a new formulation
for massive superspin-$s$ multiplet in AdS${}_3$. 

In the Minkowski superspace limit, the dynamical systems \eqref{8.4a} and \eqref{8.4b}
are equivalent, since they are related to each other by the superfield 
Legendre transformation described in section \ref{section5}.
On the mass shell, dynamics can be recast in terms of the gauge-invariant field strength
$W_{\a(2s)}$ which obeys the equations \cite{KT}
\bea
D^\b W_{\b \a_1 \cdots \a_{2s-1}} = 0 ~ , 
\qquad
-\frac{\ri}{2} D^2  W_{\a (2s)} = m \s W_{\a (2s)}~, 
\qquad \s =\pm 1~,
\label{helicity}
\eea
 where the mass $m$ and helicity parameter $\s$ are determined 
 by $\m$.\footnote{The equations \eqref{helicity} describe the irreducible 
 massive multiplet of superhelicity $\k= (s + \frac 14 )\s$ \cite{KNT-M},
 with the $\cN=1$ superhelicity operator being defined according to \cite{MT}.}
 It is an interesting open problem to understand whether 
  the AdS models \eqref{8.4a} and \eqref{8.4b} lead to equivalent dynamics, 
  modulo a redefinition of the mass parameter $\m$. 

There are two off-shell gauge-invariant formulations for a topologically 
massive superspin-$(s+\hf)$ multiplet in AdS${}_3$. The corresponding actions are:
\begin{subequations} \label{8.5}
\bea
S^{\parallel}_{(s+\hf)}[H_{\a(2s+1)} ,L_{\a(2s-2)} |\m]
&=& {S}_{\rm{SCS}}^{(s+\hf)} [ H_{\a(2s+1)}] 
+\m^{2s-1}S^{\parallel}_{(s)}[H_{\a(2s+1)} ,L_{\a(2s-2)} ]~, ~~~\label{8.5a} \\
S^{\perp}_{(s+\hf)}[H_{\a(2s+1)} ,\U_{\b; \a(2s-2)}|\m]
&=& {S}_{\rm{SCS}}^{(s+\hf)} [ H_{\a(2s+1)}] 
\non \\ &&\qquad \qquad
+\m^{2s-1}S^{\perp}_{(s+\hf )}[H_{\a(2s+1)} , \U_{\b; \a(2s-2)}]~.
 \label{8.5b}
\eea
\end{subequations}
The theory defined by  \eqref{8.5a} was introduced in \cite{KP}, while its flat-superspace 
counterpart appeared earlier in \cite{KT}. The other model, eq. \eqref{8.5b}, is a new formulation for a massive superspin-$(s+\hf) $ multiplet in AdS${}_3$. 

In the Minkowski superspace limit, the dynamical systems \eqref{8.5a} and \eqref{8.5b}
are equivalent, since they are related to each other by the superfield 
Legendre transformation described in section \ref{section5}.
 It is also an interesting open problem to understand whether 
  the models \eqref{8.5a} and 
 \eqref{8.5b} in AdS${}_3$ generate equivalent dynamics.

We now present two off-shell formulations for the massive $\cN=1$ gravitino supermultiplet  in  AdS${}_3$  and analyse the corresponding 
equations of motion.\footnote{The
construction of the models \eqref{massL} and \eqref{massT}
is similar to those used to derive the off-shell formulations for massive superspin-1
and superspin-3/2 multiplets in four dimensions 
\cite{OS2,BL1,BL2,AB1,AB2,BGLP1,BGLP2,GSS,BGKP,Gates:2005su,GKT-M}.}
The massive extension of the longitudinal theory \eqref{425} is described by the action
\bea
S^{||}_{(1), \,\m}
&=& -\frac{1}{2}
\int 
\rd^{3|2}z
\, E \,\bigg\{
\frac{\ri}{2} H^{\a \b} {\de}^2 H_{\a \b}
-\frac{\ri}{2} \de_{\b}H^{\a \b} \nabla^{\g} H_{\g \a}  - V \nabla^{\a \b} H_{\a \b}
\non \\
&&
+\frac{\ri}{2} V \nabla^2 V + (\m + 2 \cS) H^{\a \b} H_{\a \b} -2( \m -2 \cS) V^2
\bigg\}~,
\label{massL}
\eea
with $\m$ a real mass parameter. 
The massive gravitino action is thus constructed from the massless one by adding mass-like terms. In the limit $\m\to 0$, the action reduces to \eqref{425}.
The equations of motion for the dynamical superfields $H^{\a \b}$ and $V$ are
\bsubeq
\bea
&& \qquad 2 \de^{\g}\,_{(\a}H_{\b) \g} -\ri \de^2 H_{\a \b}-2 \de_{\a \b}V - 4 \m H_{\a \b} = 0~, \label{m1}\\
&& \qquad \de^{\a \b} H_{\a \b} = \big(\ri \de^2 + 8\cS -4\m \big)V~.
\label{m2}
\eea
\esubeq
Multiplying \eqref{m1} by $\de^{\a \b}$ and noting that $[\de_{\a \b}, \de^2]=0$ yields
\bea
-\ri \de^2 \de^{\a \b}H_{\a \b}+ 4 \Box V -4 \m \de^{\a \b} H_{\a \b}=0~. \label{m3}
\eea
Substituting \eqref{m2} into \eqref{m3} leads to
\bea
V=0~.
\eea
Now that $V=0$ on-shell, eq.~\eqref{m2} turns into
\bea
\de^{\a \b}H_{\a \b}=0~,
\eea
while \eqref{m1} can equivalently be written as
\bea
-\ri \de^{\g} \de_{\a}H_{\b \g}-(2 \m + 4 \cS) H_{\a \b}=0~. \label{m4}
\eea
Making use of the identity \eqref{A2},
it immediately follows from \eqref{m4} that
\bea
\de^{\a} H_{\a \b}=0~,
\eea
and then \eqref{m4} is equivalent to
\bea
-\frac{\ri}{2}\de^2 H_{\a \b} = (\m + 2\cS) H_{\a \b}~.
\eea
Therefore, we have demonstrated that the model \eqref{massL} leads to the following conditions on the mass shell:
\bsubeq \label{8.15}
\bea
V &=& 0~,
\\
\de^{\a}H_{\a \b}&=&0 \quad \implies \quad \de^{\a \b}H_{\a \b}=0~, \\
-\frac{\ri}{2}\de^2 H_{\a \b} &=& (\m +2\cS ) H_{\a \b}~.
\eea
\esubeq
Such conditions are required to describe an irreducible on-shell massive gravitino multiplet in 3D ${\cal N}=1$ AdS superspace \cite{KNT-M}. 

In the transverse formulation \eqref{337}, the action for a massive gravitino multiplet is defined by
\bea
S^{\perp}_{(1),\, \m}
&=&
- \hf \int 
\rd^{3|2}z 
\, E \, \bigg\{\frac{\ri}{2} H^{\a \b} {\de}^2 H_{\a \b}
-{\ri} \de_{\b}H^{\a \b} \nabla^{\g} H_{\g \a} 
-H^{\a \b} {\de}_{\a} \cW_{\b}
- \frac{\ri}{2} \cW^{\a}\cW_{\a} \non\\
&&+(\m + 4\cS) H^{\a \b}H_{\a \b} 
- \ri (\mu + 2\cS) \Big(\J^{\a}\cW_{\a} 
+2 \m \J^\a\J_\a \Big)
\bigg\}~.
\label{massT}
\eea
In the limit $\m\to 0$, the action reduces to \eqref{337}.
One may check that the equations of motion for this model imply that 
\bsubeq \label{8.17}
\bea
\J_{\a} &=& 0~,
\\
\de^{\a}H_{\a \b}&=&0 \quad \implies \quad \de^{\a \b}H_{\a \b}=0~, \\
-\frac{\ri}{2}\de^2 H_{\a \b} &=& (\m +4\cS ) H_{\a \b}~.
\eea
\esubeq

The actions \eqref{massL} and \eqref{massT} can be made into gauge-invariant ones using the Stueckelberg construction. 

In the Minkowski superspace limit, the massive models \eqref{massL} and \eqref{massT} 
lead to the identical equations of motion described in terms of $H_{\a\b}$:
\bea
D^{\a}H_{\a \b}=0 ~,\qquad 
-\frac{\ri}{2}D^2 H_{\a \b} &=& \m H_{\a \b}~.
\eea
In the AdS case, the equations \eqref{8.15} and \eqref{8.17} lead to equivalent 
dynamics modulo a redefinition of $\m$.
It is an interesting open problem to understand whether there exists a duality 
 transformation relating these models. 

There exist  alternative off-shell gauge-invariant formulations 
for massive higher-spin supermultiplets in AdS${}_3$ proposed in \cite{KP} 
for $\cN=1$ AdS supersymmetry and in \cite{HK18} for (2,0) AdS supersymmetry.
In the $\cN=1$ case the corresponding action is 
\bea
{S}_{\rm{massive}}^{(n/2)} [ H_{\a(n)}] 
= - \frac{\ri^n}{2^{\left \lfloor{n/2}\right \rfloor +1} \m}
   \int \rd^{3|2}z \, E\,
 {H}^{\a(n)} \Big( \m +\frac{\ri}{2} \nabla^2 \Big)  
{W}_{\a(n) }( {H}) ~, 
\label{8.19}
\eea
with $\m \neq 0$ a real parameter.
This action may be viewed as a deformation of the superconformal model \eqref{8.1}.
It is invariant under the gauge transformation \eqref{3.299}
as a consequence of the condition \eqref{8.2b} and the identity \eqref{A3}.

In the flat superspace limit,
the action \eqref{8.19} leads to the equation of motion 
\bea
-\frac{\ri}{2} D^2
{W}_{\a(n) } = \m W_{\a(n)} ~.
\eea
Since $W_{\a(n)}$ is transverse, the equation of motion implies that  $W_{\a(n)}$ describes a massive higher-spin supermultiplet, compare with 
\eqref{helicity}.
The (2,0) supersymmetric extension of the model \eqref{8.19} 
is presented in \cite{HK18}.

It should be pointed out that there also exists  an on-shell construction 
of gauge-invariant Lagrangian formulations for massive higher-spin supermultiplets in 
${\mathbb R}^{2,1}$ and AdS${}_3$, which were  developed in \cite{BSZ3,BSZ4}.
It is obtained by combining the massive bosonic 
and fermionic higher-spin actions \cite{BSZ1,BSZ2}, and therefore 
this construction is intrinsically on-shell.
The formulations given in \cite{BSZ1,BSZ2,BSZ3,BSZ4}
are based on the gauge-invariant approaches 
to  the dynamics of massive higher-spin fields, which were advocated by Zinoviev \cite{Zinoviev} and Metsaev \cite{Metsaev}.
It is an interesting open problem to understand whether there exists an off-shell 
uplift of these models. 

All off-shell higher-spin $\cN=2$ supermultiplets in AdS${}_3$,
both with (2,0) and (1,1) AdS supersymmetry \cite{HK18,HKO},
are reducible gauge theories (in the terminology of the Batalin-Vilkovisky 
quantisation \cite{BV}), similar to the massless higher-spin supermultiplets 
in AdS${}_4$ \cite{KS94}. The Lagrangian quantisation of such theories is nontrivial.
In the four-dimensional case, the quantisation of the theories proposed in \cite{KS94} 
was carried out in \cite{BKS}. All off-shell higher-spin $\cN=1$ supermultiplets 
in AdS${}_3$, which we have constructed in this paper, are irreducible gauge theories
that can be quantised using the Faddeev-Popov procedure \cite{FP}
as in the non-supersymmetric case, see e.g. \cite{FH,GGS}. This opens the possibility to develop heat kernel techniques
for higher-spin theories in AdS${}^{3|2}$, as an extension of the four-dimensional 
results \cite{BK,McA2,BK86}.
\\

\noindent
{\bf Acknowledgements:}\\
We are very grateful to Dmitri Sorokin for fruitful discussions and to Darren Grasso for comments on the manuscript.
The work of JH is supported by an Australian Government Research 
Training Program (RTP) Scholarship.
The work of SMK is supported in part by the Australian 
Research Council, project No. DP160103633.

%%%%%%%%%%%%%%%%%%%%%%%%%%%%%%%%%
%%%%%%%%%%%%%%%%%%%%%%%%%%%%%%%%%

\appendix

\section{Notation, conventions and ${\cal N}=1$ AdS identities} 
\label{AppendixA}

We summarise our notation and conventions which follow \cite{KLT-M11}. The Minkowski metric is
$\eta_{ab}=\mbox{diag}(-1,1,1)$.
The spinor indices are raised and lowered by the rule
\bea
\psi^{\a}=\ve^{\a\b}\psi_\b~, \qquad \psi_{\a}=\ve_{\a\b}\psi^\b~.
\label{AA2}
\eea
Here the antisymmetric $\rm SL(2,{\mathbb R})$ invariant tensors $\ve_{\a \b}= -\ve_{\b \a}$ and $\ve^{\a \b}= -\ve^{\b \a}$ are normalised as $\ve_{12} = -1~, \ve^{12}=1$~.

We make use of real Dirac $\g$-matrices,  $\g_a := \big( (\g_a)_\a{}^\b \big)$ defined by
\bea
(\g_a)_\a{}^\b := \ve^{\b \g} (\g_a)_{\a \g} = (-\ri \s_2, \s_3, \s_1)~.
\eea
They obey the algebra
\be
\gamma_a \gamma_b=\eta_{ab}{\mathbbm{1}} + \varepsilon_{abc}
\gamma^c~,
\ee
where the Levi-Civita tensor is normalised as
$\varepsilon^{012}=-\varepsilon_{012}=1$. 
Some useful relations involving $\g$-matrices are 
\bsubeq
\bea
(\gamma^a)_{\alpha\beta}(\gamma_a)^{\rho\sigma}
&=&-(\delta_\alpha^\rho\delta_\beta^\sigma
+\delta_\alpha^\sigma\delta_\beta^\rho)~, \\
\ve_{abc}(\g^b)_{\a\b}(\g^c)_{\g\d}&=&
\ve_{\g(\a}(\g_a)_{\b)\d}
+\ve_{\d(\a}(\g_a)_{\b)\g}
~,
\\
\tr[\g_a\g_b\g_{c}\g_d]&=&
2\eta_{ab}\eta_{cd}
-2\eta_{ac}\eta_{db}
+2\eta_{ad}\eta_{bc}
~.
\eea
\esubeq

Given a three-vector $A_a$,
it can equivalently be described as a symmetric rank-2 spinor $A_{\a\b} = A_{\b \a}$,
\bea
A_{\a\b}:=(\g^a)_{\a\b}A_a~,\qquad
A_a=-\hf(\g_a)^{\a\b}A_{\a\b}~.
\eea

The relationship between the Lorentz generators with two vector indices ($M_{ab} =-M_{ba}$),  one vector index ($M_a$)
and two spinor indices ($M_{\a\b} =M_{\b\a}$) is as follows:
$M_a=\hf \ve_{abc}M^{bc}$ and $M_{\a\b}=(\g^a)_{\a\b}M_a$.
These generators 
act on a vector $V_c$ 
and a spinor $\J_\g$ 
by the rules:
\bea
M_{ab}V_c=2\eta_{c[a}V_{b]}~, ~~~~~~
M_{\a\b}\J_{\g}
=\ve_{\g(\a}\J_{\b)}~.
\label{generators}
\eea

We collect some useful identities for ${\cal N}=1$ AdS covariant derivatives, which we denote by 
$\nabla_{A} = \left( \nabla_a, \nabla_{\a} \right)$. Making use of the (anti)-commutation relation \eqref{2_0-alg-AdS-1} and \eqref{2_0-alg-AdS-2}, we obtain the following identities
\begin{subequations}
\bea
\nabla_{\a} \nabla_{\b} &=& \hf \ve_{\a \b} \nabla^2 + \ri \nabla_{\a \b} - 2 \ri \cS M_{\a \b}~, \label{A1} \\
\nabla^{\b} \nabla_{\a} \nabla_{\b} &=& 4\ri \cS \nabla_{\a}~, \label{A2} \\
\nabla^2 \nabla_{\a} &=& -\nabla_{\a} \nabla^2 + 4 \ri \cS \nabla_{\a} = 2 \ri \nabla_{\a \b} \nabla^{\b} +2 \ri \cS \nabla_{\a} - 4 \ri \cS \nabla^{\b} M_{\a \b}~, \label{A3} \\
-\frac{1}{4} \nabla^2 \nabla^2 &=& \Box -2 \ri \cS \nabla^2 + 2\cS \nabla^{\a \b} M_{\a \b} -2 \cS^2 M^{\a \b} M_{\a \b}~, \label{A4}
\eea
\end{subequations}
where $\nabla^2 = \nabla^{\a} \nabla_{\a}$ and $\Box = \nabla^{a} \nabla_{a} = -\hf \nabla^{\a \b} \nabla_{\a \b}$~. An important corollary of \eqref{A1} and \eqref{A3} is 
\bea
{[} \nabla_{\a} \nabla_{\b}, \nabla^2 {]} =0 
\quad \implies \quad {[} \nabla_{\a \b}, \nabla^2 {]}=0~.
\eea
The left-hand side of \eqref{A4} can be expressed in terms of the quadratic Casimir operator of the 3D ${\cal N}=1$ AdS supergroup \cite{KP}:
\bea
\mathbb{Q} = -\frac{1}{4} \nabla^2 \nabla^2 + \ri \cS \nabla^2~, \qquad {[} \mathbb{Q}, \nabla_A {]} =0~.
\label{casimir}
\eea
We also note the following commutation relation
\bea
{[}(\bm \de^{\1})^2 (\bm \de^{\1})^2 -4 \ri \cS (\bm \de^{\1})^2, \bm \nabla_{\a}^{\2} {]} &=& 16 \cS \nabla_{\a \b} \bm \nabla^{\2 \b} -16 \cS^2 \bm \nabla_{\a}^{\2} \non\\
&-&32 \cS^2 \bm \nabla^{\2 \b} M_{\a \b} -32 \ri \cS^2 \bm \nabla^{\1}_{\a} J~.
\eea

Given an arbitrary superfield $F$ and its complex conjugate $\bar F $, the following relation holds 
\bea
\overline{\nabla_{\a} F} = - (-1)^{\e(F)} \nabla_{\a} \bar F~,
\eea
where $\e(F)$ denotes the Grassmann parity of $F$~.

\section{Component structure of ${\cal N}=1$ higher-spin actions} 
\label{AppendixB}
In this appendix we will discuss the component structure of the two new off-shell ${\cal N}=1$ supersymmetric higher-spin theories: the transverse massless superspin-$s$ multiplet \eqref{action-t2-new}, and the transverse massless superspin-$(s+\hf)$ multiplet \eqref{action-t3-new}. For simplicity we will carry out our analysis in flat Minkowski superspace. 
In accordance with \eqref{252}, the component form of an ${\cal N}=1$ supersymmetric action is computed by the rule
\bea
S= \int \rd^{3|2}z
 \, L = \frac{\ri}{4} \int \rd^3 x \, D^2 L \Big|_{\q=0}~, \qquad L = \bar{L}~. 
\label{comp}
\eea

%%%%%%%%%%%%%%%%%%%%%%%%%%%%%%
%%%%%%%%%%%%%%%%%%%%%%%%%%%%%%%

\subsection{Massless superspin-$s$ action}
Let us first work out the component structure of the massless integer superspin model \eqref{action-t2-new}. In the flat-superspace limit, the transverse action \eqref{action-t2-new} takes the form
\bea
\lefteqn{S^{\perp}_{(s)}[H_{\a(2s)} ,{\Psi}_{\b; \,\a(2s-2)} ]
= \Big(-\hf \Big)^{s} 
\int 
\rd^{3|2}z\,
 \bigg\{\frac{\ri}{2} H^{\a(2s)} D^2 H_{\a(2s)}}
\non \\
&&\qquad - \ri s D_{\b} H^{\b \a(2s-1)} D^{\g}H_{\g \a(2s-1)}  -(2s-1) \cW^{\b \a(2s-2)} D^{\g} H_{\g \b \a(2s-2)} 
\non \\
&& \qquad -\frac{\ri}{2} (2s-1)\Big(\cW^{\b ;\, \a(2s-2)} \cW_{\b ;\, \a(2s-2)}+\frac{s-1}{s} \cW_{\b;}\,^{\b \a(2s-3)} \cW^{\g ;}\,_{\g \a(2s-3)} \Big) 
\bigg\}~.
\eea

As described in \eqref{3.355}, it is possible to choose a gauge condition $\J_{(\a_1;\, \a_2 \cdots \a_{2s-1})}=0$, such that the above action turns into 
\bea
\lefteqn{S^{\perp}_{(s)}[H_{\a(2s)} ,{\Psi}_{\b; \,\a(2s-2)} ]
= \Big(-\hf \Big)^{s} 
\int 
\rd^{3|2}z\,
 \bigg\{\frac{\ri}{2} H^{\a(2s)} D^2 H_{\a(2s)}}
\non \\
&& \qquad \quad - \ri s D_{\b} H^{\b \a(2s-1)} D^{\g}H_{\g \a(2s-1)} -2(s-1) \vf^{\a(2s-3)} \pa^{\b \g} D^{\d} H_{\b \g \d \a(2s-3)}
\non \\
&&\qquad \quad -\frac{2\ri}{s}(s-1) \vf^{\a(2s-3)} \Box \vf_{\a(2s-3)} -\frac{\ri (s-1)(s-2)(2s-3)}{s(2s-1)} \pa_{\d \l}\vf^{\d \l \a(2s-5)} \pa^{\b \g} \vf_{\b \g \a(2s-5)} \non \\
&&\qquad \quad + \frac{\ri(s-1)(2s-3)}{2s(2s-1)} D_{\b} \vf^{\b \a(2s-4)} D^2 D^{\g} \vf_{\g \a(2s-4)}
\bigg\}~.
\label{B.1}
\eea
It is invariant under the following gauge transformations
\begin{subequations}
\bea
\d H_{\a(2s)} &=& -\pa_{(\a_1 \a_2} \eta_{\a_3 \dots \a_{2s})}~,\\
\d \vf_{\a(2s-3)}&=& \ri D^{\b} \eta_{\b \a(2s-3)}~,
\eea
\label{B.2}
\end{subequations}
where the gauge parameter $\eta_{\a(2s-2)}$ is a real unconstrained superfield. 

The gauge freedom \eqref{B.2} can be used to impose a Wess-Zumino gauge
\bea
\vf_{\a(2s-3)}|=0~, \qquad D_{(\a_1} \vf_{\a_2 \cdots \a_{2s-2})}| =0~. 
\label{B.3}
\eea
In order to preserve these gauge conditions, the residual gauge freedom has to be constrained by
\bea
D^{\b}\eta_{\b \a(2s-3)}|=0~, \qquad D^2 \eta_{\a(2s-2)}| = 2\ri \, \pa^{\b}\,_{(\a_1} \eta_{\a_2 \cdots \a_{2s-2}) \b}|~.
\eea
These imply that there remain two independent, real components of $\eta_{\a(2s-2)}$:
\bea
\xi_{\a(2s-2)}:= \eta_{\a(2s-2)}|~, \qquad \l_{\a(2s-1)} := \ri D_{(\a_1} \eta_{\a_2 \cdots \a_{2s-1})}|~.
\eea
In the gauge \eqref{B.3}, the independent component fields of $\vf_{\a(2s-3)}$ can be chosen as 
\bea
y_{\a(2s-4)}:= -\frac{2s-2}{2s-1} D^{\b} \vf_{\b \a_1 \cdots \a_{2s-4}}|~, \qquad y_{\a(2s-3)} := \frac{\ri}{2} D^2 \vf_{\a(2s-3)}|~.
\eea
We define the component fields of $H_{\a(2s)}$ as
\bea
h_{\a(2s)} &:=& -H_{\a(2s)}|~,\\
h_{\a(2s+1)} &:=& \ri \frac{s}{2s+1} D_{(\a_1} H_{\a_2 \cdots \a_{2s+1})}|~, \qquad 
y_{\a(2s-1)} := \ri D^{\b} H_{\b \a_1 \cdots \a_{2s-1}}|~,\\
F_{\a(2s)} &:=& \frac{\ri}{4} D^2 H_{\a(2s)} |~.
\eea

Applying the reduction rule \eqref{comp} to the ${\cal N}=1$ action \eqref{B.1}, we find that it splits into bosonic and fermionic parts:
\bea
S^{\perp}_{(s)}[H_{\a(2s)} ,{\Psi}_{\b; \,\a(2s-2)} ] = S_{\rm bos} + S_{\rm ferm}~.
\eea
The bosonic action is given by
\bea
S_{\rm bos}
&=& \Big(-\hf \Big)^{s}
\int 
\rd^3x \,
 \bigg\{ 2(1-s) F^{\a(2s)} F_{\a(2s)} + 2s F^{\a(2s-1) \b} \pa^{\g}\, _{\b} h_{\a(2s-1)\g}
\non\\
&&-\hf (s-1) h^{\a(2s)} \Box h_{\a(2s)}-\frac{(2s-1)(2s-3)}{2s(s-1)}y^{\a(2s-4)} \Box y_{\a(2s-4)} 
\non\\
&&- \frac{(2s-1)(2s-3)}{4(s-1)} y^{\a(2s-4)}\pa^{\b\g}\pa^{\d \l}h_{\b \g \d \l \a(2s-4)}
\non\\
&&- \frac{(s-2)(2s-1)(2s-3)(2s-5)}{16s(s-1)^2} \pa_{\d \l} y^{\d \l \a(2s-6)} \pa^{\b \g} y_{\b \g \a(2s-6)}
\bigg\}~.
\eea
Integrating out the auxiliary field $F_{\a(2s-2)}$ leads to 
\bea
S_{\rm bos}
&=& \Big(-\hf \Big)^{s} \,\frac{2s-1}{2s-2}
\int 
\rd^3x \,
 \bigg\{h^{\a(2s)} \Box h_{\a(2s)}-\frac{s}{2}\pa_{\d \l} h^{\d \l \a(2s-2)} \pa^{\b \g} h_{\b \g \a(2s-2)}
\non\\ 
&&-\frac{2s-3}{2s} \Big[ s y^{\a(2s-4)}\pa^{\b\g}\pa^{\d \l}h_{\b \g \d \l \a(2s-4)}+ 2 y^{\a(2s-4)} \Box y_{\a(2s-4)} 
\non\\
&&+ \frac{(s-2)(2s-5)}{4(s-1)} \pa_{\d \l} y^{\d \l \a(2s-6)} \pa^{\b \g} y_{\b \g \a(2s-6)} \Big]
\bigg\}~.
\label{sbos}
\eea
This action is invariant under the gauge transformations
\bea
\d_{\x} h_{\a(2s)} &=& \pa_{(\a_1 \a_2} \x_{\a_3 \cdots \a_{2s})}~,\\
\d_{\x} y_{\a(2s-4)} &=& \frac{2s-2}{2s-1} \pa^{\b\g} \x_{\b \g \a_1 \cdots \a_{2s-4}}~.
\eea
The gauge transformations for the fields $h_{\a(2s)}$ and $y_{\a(2s-4)}$ can be easily read off from the gauge transformations of the superfields $H_{\a(2s)}$ and $\vf_{\a(2s-3)}$~, respectively. 
Modulo an overall normalisation factor, \eqref{sbos} corresponds to the massless Fronsdal spin-$s$ action $S^{(2s)}_F$ described in \cite{KP}.

The fermionic sector of the component action is described by the real dynamical fields $h_{\a(2s+1)}$, $y_{\a(2s-1)}$, $y_{\a(2s-3)}$~, defined modulo gauge transformations of the form
\bea
\d_{\l} h_{\a(2s+1)} &=& \pa_{(\a_1 \a_2} \l_{\a_3 \cdots \a_{2s+1})}~,\\
\d_{\l} y_{\a(2s-1)} &=&\frac{1}{2s+1} \pa^{\b}\,_{(\a_1} \l_{\a_2 \cdots \a_{2s-1})\b}~, \\
\d_{\l} y_{\a(2s-3)} &=& \pa^{\b\g} \l_{\b \g \a_1 \cdots \a_{2s-3}}~.
\eea
The gauge-invariant action is
\bea
S_{\rm ferm}
&=& \Big(-\hf \Big)^{s} \,\frac{\ri}{2}
\int 
\rd^3x \,
 \bigg\{h^{\a(2s)\b} \pa_{\b }\,^{\g} h_{\a(2s)\g}+ 2(2s-1) y^{\a(2s-1)} \pa^{\b \g } h_{\b \g \a(2s-1)}
\non\\ 
&&+ 4(2s-1) y^{\a(2s-2) \b}\pa_{\b }\,^{\g} y_{\a(2s-2)\g} + \frac{2}{s}(2s+1)(s-1) y^{\a(2s-3)} \pa^{\b \g } y_{\b \g \a(2s-3)}
\non\\
&&- \frac{(s-1)(2s-3)}{s(2s-1)} y^{\a(2s-4)\b} \pa_{\b }\,^{\g} y_{\a(2s-4)\g}
\bigg\}~.
\label{ff-action}
\eea
It may be shown that $S_{\rm ferm}$ coincides with the Fang-Fronsdal spin-$(s+\hf)$ action, $S^{(2s+1)}_{FF}$ \cite{KP}. 

We have thus proved that at the component level and upon elimination of the auxiliary field, the transverse theory \eqref{B.1} is equivalent to a sum of two massless models: the bosonic Fronsdal spin-$s$ model and the fermionic Fang-Fronsdal spin-$(s+\hf)$ model. 

\subsection{Massless superspin-$(s+\hf)$ action}
We will now elaborate on the component structure of the massless half-integer superspin model in the transverse formulation \eqref{action-t3-new}. The theory is described in terms of the real unconstrained prepotentials $H_{\a(2s+1)}$ and $\U_{\b; \,\a(2s-2)}$. 
In Minkowski superspace, the action \eqref{action-t3-new} simplifies into
\bea
\lefteqn{S^{\perp}_{(s+\hf)}[{H}_{(2s+1)} ,\U_{\b; \,\a(2s-2)} ]
= \Big(-\hf \Big)^{s} 
\int 
\rd^{3|2}z\,
\bigg\{-\frac{\ri}{2} H^{\a(2s+1)} {\Box} H_{\a(2s+1)} } 
\non \\
&&\qquad -\frac{\ri}{8} D_{\b} H^{\b \a(2s)} D^2 D^{\g}H_{\g \a(2s)}+\frac{\ri}{8}{\pa}_{\b \g}H^{\b \g \a(2s-1)} {\pa}^{\rho \d}H_{\rho \d \a(2s-1)}
\non \\
&&\qquad -\frac{\ri}{4} (2s-1) \O^{\b; \, \a(2s-2)} \pa^{\g \d}H_{\b \g \d \a(2s-2)}
\non \\
&&\qquad  -\frac{\ri}{8}(2s-1)\Big(\O^{\b ;\, \a(2s-2)} \O_{\b ;\, \a(2s-2)}
-2(s-1)\O_{\b;}\,^{\b \a(2s-3)} \O^{\g ;}\,_{\g \a(2s-3)}  \Big) 
\bigg\}~,
\label{C.1}
\eea
with the following gauge symmetry
\begin{subequations}
\bea
\d H_{\a(2s+1)} &=& \ri D_{(\a_1} \z_{\a_2 \dots \a_{2s+1})}~,\\
\d \U_{\b;\, \a(2s-2)}&=& \frac{\ri}{2s+1}\left( D^{\g} \z_{\g \b \a(2s-2)}+ (2s+1) D_{\b} \eta_{\a(2s-2)} \right)~.
\eea \label{C1}
\end{subequations}
The action \eqref{C.1} involves the real field strength $\O_{\b;\, \a(2s-2)}$
\bea
 \O_{\b; \a(2s-2)} = -\ri D^{\g} D_{\b}\U_{\g;\a(2s-2)}~, \qquad D^{\b}\O_{\b;\, \a(2s-2)} =0~.
\eea

The gauge transformations \eqref{C1} allow us to impose a Wess-Zumino gauge on the prepotentials:
\bea
H_{\a(2s+1)}|=0~, \quad D^{\b}H_{\b \a_1 \cdots \a_{2s}}|=0~, \quad \U_{\b; \, \a(2s-2)}|=0~, \quad D^{\b} \U_{\b; \, \a(2s-2)}|=0~.
\label{C.2}
\eea
The residual gauge symmetry preserving the gauge conditions \eqref{C.2} is characterised by
\bsubeq
\bea
D_{(\a_1} \z_{\a_2 \cdots \a_{2s+1})}|&=&0~, \qquad 
D^2 \z_{\a(2s)}| = -\frac{2\ri s}{s+1} \pa^{\b}\,_{(\a_1} \z_{\a_2 \cdots \a_{2s}) \b}|~, \\
D_{\b} \eta_{\a(2s-2)}| &=& D_{(\b} \eta_{\a(2s-2))}| = -\frac{1}{2s+1} D^{\g} \z_{\g \b \a(2s-2)}|~, \\
D^2 \eta_{\a(2s-2)}| &=&-\frac{\ri}{2s+1} \pa^{\b \g} \z_{\b \g \a(2s-2)}|~.
\eea
\esubeq
As a result, there are three independent, real gauge parameters at the component level, which we define as
\bea
\xi_{\a(2s)}:= \z_{\a(2s)}|~, \quad \l_{\a(2s-1)} := -\ri \frac{s}{2s+1} D^{\b} \z_{\b \a(2s-1)}|~, \quad \rho_{\a(2s-2)}:= \eta_{\a(2s-2)}|~.
\eea
Let us now represent the prepotential $\U_{\b; \, \a(2s-2)}$ in terms of its irreducible components,
\bea
\U_{\b; \,\a(2s-2)} = Y_{\b \a_1 \dots \a_{2s-2}} + \sum_{k=1}^{2s-2}\ve_{\b \a_k} Z_{\a_1 \dots \hat{\a}_k \dots \a_{2s-2}}~,
\eea
where we have introduced the two irreducible components of $\U_{\b; \, \a(2s-2)}$ by the rule
\bea
Y_{\b \a_1 \cdots \a_{2s-2}}:= \U_{(\b; \, \a_1 \cdots \a_{2s-2})}~, \qquad Z_{\a_1 \dots \a_{2s-3}} := \frac{1}{2s-1} \U^{\b;}\,_{\b \a_1 \dots \a_{2s-3}}~. 
\eea 
The next step is to determine the remaining independent component fields of $H_{\a(2s+1)}$ and $\U_{\b; \, \a(2s-2)}$ in the Wess-Zumino gauge \eqref{C.2}.

In the bosonic sector, we have the following set of fields:
\bsubeq
\bea
h_{\a(2s+2)} &:=& -D_{(\a_1}H_{\a_2 \cdots \a_{2s+2})}|~,\\
y_{\a(2s)} &:=&  D_{(\a_1} Y_{\a_2 \cdots \a_{2s})}|~,\\
z_{\a(2s-2)} &:=& -\frac{1}{s}(2s-1) D_{(\a_1} Z_{\a_2 \cdots \a_{2s-2})}|~,\\
z_{\a(2s-4)}&:=& -(2s-1)D^{\b}Z_{\b \a(2s-4)}|~.
\eea
\esubeq
Reduction of the action \eqref{C.1} to components leads to the following bosonic action:
\bea
{S}_{\rm bos}
&=& \Big(-\hf \Big)^{s} 
\int 
\rd^3x \,
 \bigg\{-\frac{1}{4}h^{\a(2s+2)} \Box h_{\a(2s+2)}+\frac{3}{16}\pa_{\d \l} h^{\d \l \a(2s)} \pa^{\b \g} h_{\b \g \a(2s)}
\non\\ 
&&+\frac{1}{4}(2s-1)\pa_{\d \l} h^{\d \l \a(2s)} \pa^{\b}\,_{(\a_1} y_{\a_2 \cdots \a_{2s}) \b}
- \frac{1}{4} (2s-1)(s-1) z^{\a(2s-2)} \pa^{\b \g} \pa^{\d \l} h_{\b \g \d \l \a(2s-2)}
\non \\
&&- \frac{1}{4}(2s-1) y^{\a(2s)} \Box y_{\a(2s)}- \frac{1}{8}(s-2)(2s-1)\pa_{\d \l} y^{\d \l \a(2s-2)} \pa^{\b \g} y_{\b \g \a(2s-2)}
\non\\
&&-(s-1)(2s-1) z^{\a(2s)} \Box z_{\a(2s)}
\non\\
&&- \frac{1}{4}(s-1)(s+2)(2s-1)(2s-3)\pa_{\d \l} z^{\d \l \a(2s-4)} \pa^{\b \g} z_{\b \g \a(2s-4)}
\non\\
&&+(s-1)(2s-1)\pa_{\b \g} y^{\b \g \a(2s-2)} \pa^{\d}\,_{(\a_1} z_{\a_2 \cdots \a_{2s-2})\d}
\non \\
&&-\frac{s}{4} \frac{2s-3}{(s-1)(2s-1)}(4s^2-12s+11) z^{\a(2s-4)} \Box z_{\a(2s-4)}
\non\\
&&+\frac{3s}{8(s-1)(2s-1)}(s-2)(2s-3)(2s-5)\pa_{\d \l} z^{\d \l \a(2s-6)} \pa^{\b \g} z_{\b \g \a(2s-6)}
\non\\
&&+ \frac{1}{4}(s+1) (2s-3) z^{\a(2s-4)} \pa^{\b \g} \pa^{\d \l} y_{\b \g \d \l \a(2s-4)}
\non \\
&&+\hf (s-2)(2s+1)(2s-3)\pa_{\b \g} z^{\b \g \a(2s-4)} \pa^{\d}\,_{(\a_1} z_{\a_2 \cdots \a_{2s-4})\d}
\bigg\}~,
\eea
which proves to be invariant under gauge transformations of the form
\bsubeq
\bea
\d_{\x} h_{\a(2s+2)} &=& \pa_{(\a_1 \a_2} \x_{\a_3 \cdots \a_{2s+2})}~, \label{bg1a}\\
\d_{\x, \rho} y_{\a(2s)} &=&-\frac{1}{s+1} \pa^{\b}\,_{(\a_1} \x_{\a_2 \cdots \a_{2s})\b}- \pa_{(\a_1 \a_2} \rho_{\a_3 \cdots \a_{2s})}~, \label{bg2a}\\
\d_{\x, \rho} z_{\a(2s-2)} &=& \frac{1}{2s(2s+1)} \pa^{\b \g} \x_{\b \g \a(2s-2)} + \frac{1}{s} \pa^{\b}\,_{(\a_1} \rho_{\a_2 \cdots \a_{2s-2})\b}~, \label{bg3a} \\
\d_{\rho} z_{\a(2s-4)} &=& \pa^{\b \g} \rho_{\b \g \a(2s-4)}~. \label{bg4a}
\eea
\esubeq

Let us consider the fermionic sector. We find that the independent fermionic fields are:
\bsubeq
\bea
h_{\a(2s+1)} &:=& \frac{\ri}{4} D^2 H_{\a(2s+1)} |~, \\
y_{\a(2s-1)} &:=& \frac{\ri}{8} D^2 Y_{\a(2s-1)}|~,\\
y_{\a(2s-3)} &:=& \frac{\ri}{2}s(2s-1) D^2 Z_{\a(2s-3)}|~,
\eea
\esubeq
and their gauge transformation laws are given by
\bsubeq
\bea
\d_{\l} h_{\a(2s+1)} &=& \pa_{(\a_1 \a_2} \l_{\a_3 \cdots \a_{2s+2})}~, \label{fg1a}\\ 
\d_{\l} y_{\a(2s-1)} &=& \frac{1}{2s+1} \pa^{\b}\,_{(\a_1} \l_{\a_2 \cdots \a_{2s-1})\b}~, \label{fg2a}\\
\d_{\l} y_{\a(2s-3)} &=& \pa^{\b \g } \l_{\b \g \a(2s-3)}~. \label{fg3a}
\eea
\esubeq
The above fermionic fields correspond to the dynamical variables of the Fang-Fronsdal spin-$(s + \hf)$ model. As follows from \eqref{fg1a}, \eqref{fg2a} and \eqref{fg3a}, their gauge freedom is equivalent to that of the
massless spin-$(s + \hf)$ gauge field. 
Indeed, direct calculations of the component action give the standard massless gauge-invariant spin-$(s+\hf)$ action $S^{(2s+1)}_{FF}$.

The component structure of the obtained supermultiplets is a three-dimensional counterpart of so-called (reducible) higher-spin triplet systems. 
In AdS${}_D$ an action for bosonic higher-spin triplets was constructed in
\cite{Sagnotti:2003qa}
and for fermionic triplets in \cite{Sorokin:2008tf,Agugliaro:2016ngl}.
 Our superfield construction  provides a manifestly off-shell supersymmetric generalisation of these systems.
It might be of interest to extend it to AdS${}_4$.

\begin{footnotesize}

\end{footnotesize}

\end{document}